\documentclass[a4paper,final,11p,onecolumn]{revtex4}

\usepackage[breaklinks]{hyperref}
\usepackage{pstricks}
\usepackage{amsmath}
\usepackage{bm}
\usepackage{longtable}
\usepackage{multirow}
\usepackage[utf8]{inputenc}

\newcommand{\br}{{\bf r}}

\newcounter{bla}

\begin{document}

\title{{\sc Libxc}: a library of exchange and correlation functionals for density functional theory}

\author{Miguel A. L. Marques}
\email[Corresponding author. Electronic address:\;]{marques@tddft.org}
\affiliation{Universit\'e de Lyon, F-69000 Lyon, France and  LPMCN, CNRS, UMR 5586, \\Universit\'e Lyon 1, F-69622 Villeurbanne, France}

\author{Micael J. T. Oliveira}
\affiliation{Center for Computational Physics, University of Coimbra, \\Rua Larga, 3004-516 Coimbra, Portugal}

\author{Tobias Burnus}
\affiliation{Peter Gr\"unberg Institut and Institute for Advanced Simulation,\\Forschungszentrum J\"ulich, and J\"ulich Aachen Research Alliance, 52425 J\"ulich, Germany}

\begin{abstract}
  The central quantity of density functional theory is the so-called
  exchange-correlation functional. This quantity encompasses all
  non-trivial many-body effects of the ground-state and has to be
  approximated in any practical application of the theory. For the
  past 50 years, hundreds of such approximations have appeared, with
  many successfully persisting in the electronic structure community
  and literature. Here, we present a library that contains routines to
  evaluate many of these functionals (around 180) and their
  derivatives.
\end{abstract}

\maketitle

\section{Introduction}
Density functional theory (DFT) is perhaps one of the most successful
theories in Physics and in Chemistry of the last
half-century~\cite{PhysRev.136.B864,Parr1989,Dreizler90,Fiolhais2003}. It is currently
used to predict the structure and the properties of atoms, molecules,
and solids; it is a key ingredient of the new field of Materials
Design, where one tries to create new materials with specific
properties; it is making its way in Biology as an important tool in
the investigation of proteins, DNA, etc. These are only a few examples
of a discipline that even now, almost 50 years after its birth, is
growing at an exponential rate.

Almost all applications of DFT are performed within the so-called
Kohn-Sham scheme~\cite{PhysRev.140.A1133}, that uses a non-interacting
electronic system to calculate the density of the interacting
system~\cite{vanLeeuwen200325}. The Kohn-Sham scheme leeds to the
following equations. (Hartree atomic units are used throughout the
paper, i.e.  $e^2=\hbar=m_e=1$.)
\begin{equation}
  \left[ -\frac{\nabla^2}{2} + v_{\rm ext}(\br) +  v_{\rm Hartree}[n](\br) + v_{\rm xc}[n](\br)\right] \psi_i(\br)
  = \varepsilon_i \psi_i(\br)
  \,,
\end{equation}
where the first term represents the kinetic energy of the electrons;
the second is the external potential usually generated by a set of
Coulombic point charges (sometimes described by pseudopotentials); the
third term is the Hartree potential that describes the classical
electrostatic repulsion between the electrons,
\begin{equation}
 v_{\rm Hartree}[n](\br) = \int\!\!d^3r'\,\frac{n(\br')}{|\br-\br'|}
  \,,
\end{equation}
and the exchange-correlation (xc) potential $v_{\rm xc}[n]$ is defined by
\begin{equation}
  v_{\rm xc}[n](\br) = \frac{\delta E_{\rm xc}[n]}{\delta n(\br)}
  \,.
\end{equation}
$E_{\rm xc}[n]$ is the xc energy functional. Note that by $[n]$ we
denote that the quantity is a {\it functional} of the electronic
density,
\begin{equation}
  n(\br) = \sum_i^{\rm occ.} |\psi_i(\br)|^2
  \,,
\end{equation}
where the sum runs over the occupied states.

The central quantity of this scheme is the xc energy $E_{\rm xc}[n]$
that describes all non-trivial many-body effects.  Clearly, the exact
form of this quantity is unknown and it must be approximated in any
practical application of DFT. We emphasize that the precision of any
DFT calculation depends solely on the form of this quantity, as this
is the only real approximation in DFT (neglecting numerical
approximations that are normally controllable).

The first approximation to the exchange energy, the local density
approximation (LDA), was already proposed by Kohn and Sham in the same
paper where they described their Kohn-Sham
scheme~\cite{PhysRev.140.A1133}. It states that the value of xc energy
density at any point in space is simply given by the xc energy density
of a homogeneous electron gas (HEG) with electronic density $n(\br)$.
Mathematically, this is written as
\begin{equation}
  E^{\rm LDA}_{\rm xc} = \int\!\!d^3r\, n(\br) e^{\rm HEG}_{\rm xc}(n(\br))
  \,,
\end{equation}
where $e^{\rm HEG}_{\rm xc}(n)$ is the xc energy {\it per electron} of
the HEG. Note that this quantity is a {\it function} of $n$. While the
exchange contribution $e^{\rm HEG}_{\rm x}(n)$ can be easily
calculated analytically, the correlation contribution is usually taken
from Quantum Monte-Carlo
simulations~\cite{PhysRevLett.45.566,PhysRevB.50.1391}. As defined,
the LDA is unique, but, as we will see in the \ref{app:1}, even such
precise definition can give rise to many different parameterizations.

\begin{figure}[t]
  \centering
  \begin{pspicture}(12,8)
    \psline[linecolor=black](5,1)(5,7)
    \psline[linecolor=black](7,1)(7,7)
    \psline[linecolor=black](5,2)(7,2)
    \psline[linecolor=black](5,3)(7,3)
    \psline[linecolor=black](5,4)(7,4)
    \psline[linecolor=black](5,5)(7,5)
    \psline[linecolor=black](5,6)(7,6)
    \rput(6,0.25){Hartree world}
    \rput(6,7.5){Chemical accuracy}
    \rput[l](7.5,2){LDA}
    \rput[l](7.5,3){GGA}
    \rput[l](7.5,4){meta-GGA}
    \rput[l](7.5,5){EXX with correlation}
    \rput[l](7.5,6){EXX with partial exact correlation}
    \rput[r](4.5,2){$n(\br)$}
    \rput[r](4.5,3){$\nabla n(\br)$}
    \rput[r](4.5,4){$\nabla^2 n(\br), \tau(\br)$}
    \rput[r](4.5,5){$\psi_i(\br)$ (occupied)}
    \rput[r](4.5,6){$\psi_i(\br)$ (empty)}
  \end{pspicture}
  \caption{Jacob's ladder of density functional approximations for the
    xc energy.}
  \label{fig:jacob}
\end{figure}
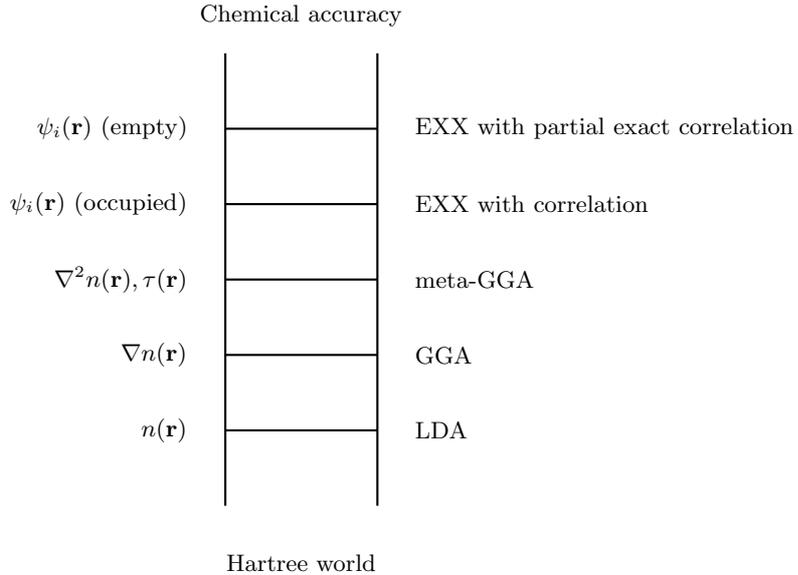

During the past 50 years, hundreds of different forms
appeared~\cite{Scuseria2005} and they are usually arranged in
families, which have names such as generalized-gradient approximations
(GGAs), meta-GGAs, hybrid functionals, etc. In 2001, John Perdew came
up with a beautiful idea on how to illustrate these families and their
relationship~\cite{perdew:1}. He ordered these families as rungs in a ladder that
leads to the heaven of ``chemical accuracy'', and that he christened
the Jacob's ladder of density functional approximations for the xc
energy (see Fig.~\ref{fig:jacob}).  Every rung adds a dependency on
another quantity, thereby increasing the precision of the functional
but also increasing the numerical complexity and the computational
time.

At the bottom of the ladder we find the LDA, a functional that depends
locally on the density only. The second rung is occupied by the GGA
\begin{equation}
  E^{\rm GGA}_{\rm xc} = \int\!\!d^3r\, n(\br) e^{\rm GGA}_{\rm xc}(n(\br), \nabla n(\br))
  \,.
\end{equation}
As one can see, one now adds the gradient of the density, a semi-local
quantity that depends on an infinitesimal region around $\br$, as a
parameter to the energy density. Note that there is a considerable
amount of craftsmanship and physical/chemical intuition going into the
creation of the function $e^{\rm GGA}_{\rm xc}(n, \nabla n)$, but also
a fair quantity of arbitrariness. It is therefore not surprising that
many different forms were proposed over the years. The same is true
for the functionals on the next rung, the meta-GGAs
\begin{equation}
  E^{\rm mGGA}_{\rm xc} = \int\!\!d^3r\, n(\br) e^{\rm mGGA}_{\rm xc}(n(\br), \nabla n(\br), \nabla^2 n(\br), \tau(\br))
  \,.
\end{equation}
This time one adds the Laplacian of the density $\nabla^2 n(\br)$ and
also (twice) the kinetic energy density
\begin{equation}
  \tau(\br) = \sum_i^{\rm occ}|\nabla \psi_i(\br)|^2
  \,.
\end{equation}
Note that the meta-GGAs are effectively orbital functionals due to the
dependence in $\tau(\br)$.

The forth rung is occupied by functionals that include the
exact-exchange (EXX) contribution to the energy
\begin{equation}
  E^{\rm EXX}_{\rm x} = -\frac{1}{2}\int\!\!d^3r\int\!\!d^3r'\,
  \frac{\psi_i(\br)\psi_i^*(\br')\psi_j(\br')\psi_j^*(\br)}{|\br - \br'|}
  \,.
\end{equation}
These functionals can include the whole $E^{\rm EXX}_{\rm x}$ or only
a fraction of it, and $E^{\rm EXX}_{\rm x}$ can be evaluated with the
bare Coulomb interaction or with a screened version of it.
Furthermore, one can add a (semi-)local xc term or one that depends
on the orbitals. In any case, the fourth rung only includes functionals
that depend on the {\it occupied} orbitals only. Notorious examples of
functionals on this rung are the hybrid functionals
\begin{equation}
  E^{\rm Hyb}_{\rm xc} = a_x E^{\rm EXX}_{\rm x} + E^{\rm mGGA}_{\rm xc}[n(\br), \nabla n(\br), \nabla^2 n(\br), \tau(\br)]
  \,.
\end{equation}
Note that the local part of this functional can be a meta-GGA, a GGA,
or even an LDA.

Finally, on the last rung of Jacob's ladder one finds functionals that
depend on the empty (virtual) Kohn-Sham orbitals. Perhaps the best
known example of these functionals is the random-phase approximation
(RPA).

In a practical DFT calculation one needs typically to evaluate both
$E_{\rm xc}$ and $v_{\rm xc}$. Furthermore, to obtain response
properties higher derivatives of $E_{\rm xc}$ are required. For
example, in first order one can get the electric polarizability, the
magnetic susceptibility, phonon frequencies, etc., and these usually
require the knowledge of the xc kernel
\begin{equation}
  f_{\rm xc}(\br, \br') = \frac{\delta E_{\rm xc}}{\delta n(\br) \delta n(\br')}
  \,.
\end{equation}
In second order one can obtain, e.g., hyperpolarizabilities, Raman
tensors, etc., but these calculations usually require
\begin{equation}
  k_{\rm xc}(\br, \br', \br'') = \frac{\delta E_{\rm xc}}{\delta n(\br) \delta n(\br') \delta n(\br'')}
  \,.
\end{equation}
The calculation of these derivatives is fairly straightforward by
using basic functional analysis and the chain rule for functional
derivatives. We give here as an example the case of the xc potential
for a GGA
\begin{multline}
  v^{\rm GGA}_{\rm xc}(\br)  = e^{\rm GGA}_{\rm xc}(n(\br), \nabla n(\br))
  + \int\!\!d^3r\, n(\br) \left.\frac{\partial e^{\rm GGA}_{\rm xc}}{\partial n}
  \right|_{\scriptsize\begin{array}{cc}n=n(\br) \\ \nabla n = \nabla n(\br) \end{array}}
  \delta(\br-\br') \\
  + \int\!\!d^3r\, n(\br) \left.\frac{\partial e^{\rm GGA}_{\rm xc}}{\partial \nabla n}
  \right|_{\scriptsize\begin{array}{cc}n=n(\br) \\ \nabla n = \nabla n(\br) \end{array}}
  \nabla \delta(\br-\br')
  \,.
\end{multline}
Integrating by parts the third term, one finally arrives at
\begin{multline}
  v^{\rm GGA}_{\rm xc}(\br)  = e^{\rm GGA}_{\rm xc}(n(\br), \nabla n(\br))
  +  n(\br) \left.\frac{\partial e^{\rm GGA}_{\rm xc}}{\partial n}
  \right|_{\scriptsize\begin{array}{cc}n=n(\br) \\ \nabla n = \nabla n(\br) \end{array}}
  - \nabla \left[n(\br) \left.\frac{\partial e^{\rm GGA}_{\rm xc}}{\partial \nabla n}
  \right|_{\scriptsize\begin{array}{cc}n=n(\br) \\ \nabla n = \nabla n(\br) \end{array}}\right]
  \,.
\end{multline}
Clearly, higher functional derivatives of $E_{\rm xc}$ involve
higher partial derivatives of $e_{\rm xc}$.

By now it is clear what a code needs to implement for the functionals of
the first three rungs: given $n(\br)$, and possibly $\nabla n(\br)$,
$\nabla^2 n(\br)$, and $\tau (\br)$, one needs the xc energy density
$e_{\rm xc}(\br)$ and all relevant partial derivatives of this
quantity. This is indeed the information that is provided by {\sc
  Libxc}. For hybrid functionals the library also returns, besides the
semi-local part, also the mixing coefficient $a_x$. Unfortunately, EXX
and other functionals of the forth and fifth rang are too dependent on
the actual numerical representation of the wave-functions and can not
be easily included in a generic library.

Finally, we would like to mention that some applications of DFT do not
use the Kohn-Sham scheme, but try to approximate the kinetic energy
functional directly in terms of the density. This approach follows the
path that was laid down in the 1927 by Thomas and
Fermi~\cite{CambridgeJournals:1732980,Fermi1927}, and is sometimes
referred to as ``orbital-free DFT''~\cite{Chen2008}. In this case, we
also need a functional form (either an LDA or a GGA) for the kinetic
energy, and many of these are present in {\sc Libxc}. A good review of
functionals for the kinetic energy density can be found in
Ref.~\cite{Ludena2002}.

Note that even if the discussion above was restricted to
spin-compensated systems for the sake of simplicity, all functionals
of {\sc Libxc} can also be used with spin-polarization.

\section{Some history}

{\sc Libxc} started as a spin-off project during the initial
development of the (time-dependent) DFT code {\sc
  Octopus}~\cite{Marques200360,PSSB:PSSB200642067,andrade:184106,Alberto:2004:1546-1955:231}.
At that point it became clear that the task of evaluation of the xc
functional was completely independent of the main structure of {\sc
  Octopus}, and could therefore be transformed into a library. The
first steps into the development of {\sc Libxc} were taken in
September 2006, and the first usable version of the library included a
few of the most popular LDA and GGA functionals.

At the same time, the European Theoretical Spectroscopy Facility was
carrying a coordinate effort to improve code interoperability and
re-usability of its software suite, so efforts were almost immediately
made to interface those codes with {\sc Libxc}. This catalyzed the
development of the library and accelerated its dissemination in the
community.

During the following years, the library was expanded following two
main lines:

(i)~To include as many functionals as possible (by now we include
around 180 --- see \ref{app:1}). This was done for several reasons.
First, we implemented nearly all of the ``old'' functionals that
played an important role in the development of DFT. From this
perspective, we can look at {\sc Libxc} as a ``living museum'' of the
history of this important discipline. It also allows users to
reproduce old results with little effort. Secondly, many of the
functionals that are proposed nowadays are implemented in {\sc Libxc}
within weeks from the moment they are published. In this way new
developments are very rapidly available in several different codes,
allowing for these new functionals to be quickly tested and
benchmarked. Note that we have an agnostic policy, i.e. we try to
include the maximum possible number of functionals, without making any
judgment of value concerning their beauty, elegance, or usefulness.
This judgment is left for the final user to perform.

(ii)~To include derivatives of the xc energy up to high orders. As we
stated before, to perform a standard Kohn-Sham calculation one only
requires the xc energy functional and its first derivatives. However,
higher derivatives are essential in order to obtain response
properties. Of course, these higher derivatives can be calculated
numerically from lower-order derivatives, but this procedure tends to
introduce unnecessary errors and instabilities in the calculations. We
therefore implemented derivatives up to the third order for the LDAs
and up to second order for the other functionals. We note that these
derivatives are hand-coded, and not automatically generated from the
output of symbolic manipulation
software~\cite{Strange2001310,JCC:JCC20758}. Even if the latter
approach is excellent in order to test implementations, the
automatically generated code is often extremely verbose, inefficient,
and unreadable. Note that {\sc Libxc} includes automatic procedures to
check the implementation of the analytic derivatives.

At the beginning, {\sc Libxc} was used exclusively in the code {\sc
  Octopus}. However, since then several other codes from both the
Solid-State Physics and Quantum Chemistry communities started to use
this library. The list of codes that use {\sc Libxc} at the time of
writing is as follows (in alphabetical order):
\begin{itemize}
  \item {\sc Abinit}~\cite{Gonze20092582,abinit2,Gonze2002478}
    (\url{http://www.abinit.org/}) --- This is a general purpose
    plane-wave code. Besides the basic functionality, {\sc Abinit}
    also includes options to optimize the geometry, to perform
    molecular dynamics simulations, or to calculate dynamical
    matrices, Born effective charges, dielectric tensors, and many
    more properties.
  \item {\sc APE}~\cite{Oliveira2008524}
    (\url{http://www.tddft.org/programs/APE}) --- The atomic
    pseudopotential engine (APE) is a tool for generating atomic
    pseudopotentials within DFT. It is distributed under the GPL and
    it produces pseudopotential files suitable for use with several
    codes.
  \item {\sc AtomPAW}~\cite{Holzwarth2001329}
    (\url{http://www.wfu.edu/~natalie/papers/pwpaw/man.html}) --- The
    computer program {\sc AtomPaw} generates projector and basis
    functions which are needed for performing electronic structure
    calculations based on the projector augmented wave (PAW)
    method. The program is applicable to materials throughout the
    periodic table.
  \item {\sc Atomistix
    ToolKit}~\cite{PhysRevB.65.165401,0953-8984-14-11-302}
    (\url{http://quantumwise.com}) --- This is a software package that
    uses non-equilibrium Green's functions simulations to study
    transport properties like I-V characteristics of nanoelectronic
    devices. It uses a powerful combination of DFT, semi-empirical
    tight-binding, and classical potentials.
  \item {\sc BigDFT}~\cite{Genovese2011149,genovese:014109}
    (\url{http://inac.cea.fr/L_Sim/BigDFT/}) --- {\sc BigDFT} is a DFT
    massively parallel electronic structure code using a wavelet basis
    set. Wavelets form a real space basis set distributed on an
    adaptive mesh. Thanks to its Poisson solver based on a Green's
    function formalism, periodic systems, surfaces and isolated
    systems can be simulated with the proper boundary conditions.
  \item {\sc DP}~\cite{QUA:QUA20486} (\url{http://www.dp-code.org/})
    --- Dielectric properties (DP) is a Linear Response TDDFT code, in
    frequency-reciprocal and frequency-real space, that uses a
    plane-wave basis set.
  \item {\sc Elk} (\url{http://elk.sourceforge.net/}) --- An
    all-electron full-potential linearized aug\-mented-plane wave
    (FP-LAPW) code with many advanced features. This code is designed
    to be as simple as possible so that new developments in the field
    of density functional theory (DFT) can be added quickly and
    reliably.  The code is freely available under the GNU General
    Public License.
  \item {\sc ERKALE} (\url{http://erkale.googlecode.com}) --- {\sc
      ERKALE} is a quantum chemistry program developed by J. Lehtola
    used to solve the electronic structure of atoms, molecules and
    molecular clusters. The main use of {\sc ERKALE} is the
    computation of X-ray properties, such as ground-state electron
    momentum densities and Compton profiles, and core (X-ray
    absorption and X-ray Raman scattering) and valence electron
    excitation spectra of atoms and molecules.
  \item {\sc Exciting}~\cite{PhysRevLett.95.136402,B903676H}
    (\url{http://exciting-code.org/}) --- {\sc Exciting} is a
    full-potential all-electron DFT package based on the linearized
    augmented plane-wave (LAPW) method. It can be applied to all kinds
    of materials, irrespective of the atomic species involved, and
    also allows for the investigation of the atomic-core region.
  \item {\sc GPAW}~\cite{PhysRevB.71.035109,0953-8984-22-25-253202}
    (\url{https://wiki.fysik.dtu.dk/gpaw}) --- {\sc GPAW} is a DFT
    Python code based on the projector-augmented wave (PAW) method and
    the atomic simulation environment (ASE). It uses real-space
    uniform grids and multigrid methods or atom-centered
    basis-functions.
  \item {\sc
      Hippo}~\cite{lathiotakis:184103,0295-5075-77-6-67003,PhysRevA.77.032509}
    --- This is an electronic structure code, developed by N.
    Lathiotakis, that implements Reduced Density Matrix Functionals
    for atomic and molecular systems using Gaussian-type orbitals .
  \item {\sc
      Octopus}~\cite{Marques200360,PSSB:PSSB200642067,andrade:184106,Alberto:2004:1546-1955:231}
    (\url{http://www.tddft.org/programs/octopus/}) --- {\sc Octopus}
    is a scientific program aimed at the {\it ab initio} virtual
    experimentation on a hopefully ever-increasing range of system
    types. Electrons are described quantum-mechanically within
    density-functional theory (DFT), and in its time-dependent form
    (TDDFT) when doing simulations in time. Nuclei are described
    classically as point particles. Electron-nucleus interaction is
    described within the pseudopotential approximation.
  \item {\sc Yambo}~\cite{Marini20091392}
    (\url{http://www.yambo-code.org/}) --- {\sc Yambo} is a Fortran/C
    code for many-body calculations in solid state and molecular
    physics. The code was originally developed in the Condensed Matter
    Theoretical Group of the Physics Department at the University of
    Rome ``Tor Vergata'' by A. Marini. Previous to its release
    under the GPL license, {\sc Yambo} was known as {\sc SELF}.
\end{itemize}

This diversity of codes is extremely important, because in this way a
certain functional can be tested and used with a variety of methods in
different physical situations. For example, a meta-GGA developed within
a certain code to get good band-gaps of solids can be immediately used
in a different code (often after a simple recompilation) to obtain the
ionization potential of molecules.

{\sc Libxc} is freely available from {\tt
  http://www.tddft.org/programs/Libxc}, and it is distributed under
the GNU Lesser General Public License v3.0. This license not only
allows everyone to read, modify, and distribute the code, but also
allows {\sc Libxc} to be linked from close-source codes. The reader is
also referred to the web site to obtain more information, updated
documentation, examples, new versions, etc.

As most open source projects, we strongly encourage contributions from
researchers willing to contribute with the implementation of new
functionals, higher derivatives, bug corrections, or even bug reports.

\section{An example}

\subsection{Calling {\sc Libxc}}

Probably the best way to explain the usage of {\sc Libxc} is through
an example. The following small program calculates the xc energy for a
given functional for several values of the density; the available C
bindings can be found in header file {\tt xc.h}. More information and
examples can be found in the manual and in the header files.

\begin{verbatim}
#include <stdio.h>
#include <xc.h>

int main()
{
  xc_func_type func;
  double rho[5] = {0.1, 0.2, 0.3, 0.4, 0.5};
  double sigma[5] = {0.2, 0.3, 0.4, 0.5, 0.6};
  double ek[5];
  int i, func_id = 1;

  /* initialize the functional */
  if(xc_func_init(&func, func_id, XC_UNPOLARIZED) != 0) {
    fprintf(stderr, "Functional '%d' not found\n", func_id);
    return 1;
  }

  /* evaluate the functional */
  switch(func.info->family)
    {
    case XC_FAMILY_LDA:
      xc_lda_exc(&func, 5, rho, ek);
      break;
    case XC_FAMILY_GGA:
    case XC_FAMILY_HYB_GGA:
      xc_gga_exc(&func, 5, rho, sigma, ek);
      break;
    }

  for(i=0; i<5; i++) {
    printf("%lf %lf\n", rho[i], ek[i]);
  }

  /* free the functional */
  xc_func_end(&func);
}
\end{verbatim}

The functionals are divided in families (LDA, GGA, etc.). Given a
functional identifier, {\tt func\_id}, the functional is
initialized by {\tt xc\_func\_init}, and evaluated by {\tt
  xc\_XXX\_exc}, which returns the energy per unit volume ({\tt ek}).
Finally, the function {\tt xc\_func\_end} cleans up. We note that we
follow the convention used in Quantum Chemistry and, instead of
passing the full gradient of the density, we use the variable
\begin{equation}
  \sigma(\br) = \nabla n(\br) \cdot \nabla n(\br)
  \,.
\end{equation}
Converting between partial derivatives with respect to $\sigma$ and
$\nabla n$ is trivially done with the use of the chain rule. All the
quantities passed to and returned by the library are in atomic units.

Fortran 90 bindings are also included in {\sc Libxc}. These can be
found in the file {\tt libxc\_master.F90}. In general, calling {\sc
  Libxc} from Fortran is as simple as from C. Here is the previous
example in Fortran:

\begin{verbatim}
program lxctest
   use xc_f90_types_m
   use xc_f90_lib_m

   implicit none

   TYPE(xc_f90_pointer_t) :: xc_func
   TYPE(xc_f90_pointer_t) :: xc_info
   real(8) :: rho(5) = (/0.1, 0.2, 0.3, 0.4, 0.5/)
   real(8) :: sigma(5) = (/0.2, 0.3, 0.4, 0.5, 0.6/)
   real(8) :: ek(5)
   integer :: i, func_id

   func_id = 1

   ! initialize the functional
   call xc_f90_func_init(xc_func, xc_info, func_id, XC_UNPOLARIZED)

   ! evaluate the functional
   select case (xc_f90_info_family(xc_info))
   case(XC_FAMILY_LDA)
     call xc_f90_lda_exc(xc_func, 5, rho(1), ek(1))
   case(XC_FAMILY_GGA, XC_FAMILY_HYB_GGA)
     call xc_f90_gga_exc(xc_func, 5, rho(1), sigma(1), ek(1))
   end select

   do i = 1, 5
     write(*,"(F8.6,1X,F9.6)") rho(i), ek(i)
   end do

   ! free the functional
   call xc_f90_func_end(xc_func)

end program lxctest
\end{verbatim}

\subsection{The info structure}

Besides the mathematical formulas necessary to evaluate the functional
and its derivatives, {\sc Libxc} includes a considerable amount of
metadata that is quite useful for both the calling program and for the
end user. This information is contained for each functional in the
structure {\tt xc\_func\_info\_type}. The relevant part of this structure
for the end user is defined as

\begin{verbatim}
/* flags that can be used in info.flags */
#define XC_FLAGS_HAVE_EXC         (1 <<  0) /*    1 */
#define XC_FLAGS_HAVE_VXC         (1 <<  1) /*    2 */
#define XC_FLAGS_HAVE_FXC         (1 <<  2) /*    4 */
#define XC_FLAGS_HAVE_KXC         (1 <<  3) /*    8 */
#define XC_FLAGS_HAVE_LXC         (1 <<  4) /*   16 */
#define XC_FLAGS_1D               (1 <<  5) /*   32 */
#define XC_FLAGS_2D               (1 <<  6) /*   64 */
#define XC_FLAGS_3D               (1 <<  7) /*  128 */
#define XC_FLAGS_STABLE           (1 <<  9) /*  512 */
#define XC_FLAGS_DEVELOPMENT      (1 << 10) /* 1024 */

typedef struct{
  int   number;   /* indentifier number */
  int   kind;     /* XC_EXCHANGE and/or XC_CORRELATION */

  char *name;     /* name of the functional, e.g. "PBE" */
  int   family;   /* type of the functional, e.g. XC_FAMILY_GGA */
  char *refs;     /* references                       */

  int   flags;    /* see above for a list of possible flags */

  ...
} xc_func_info_type;
\end{verbatim}

For example, for the Slater exchange functional, this structure is defined as

\begin{verbatim}
const XC(func_info_type) XC(func_info_lda_x) = {
  XC_LDA_X,
  XC_EXCHANGE,
  "Slater exchange",
  XC_FAMILY_LDA,
  "PAM Dirac, Proceedings of the 
    Cambridge Philosophical Society 26, 376 (1930)\n"
  "F Bloch, Zeitschrift fuer Physik 57, 545 (1929)",
  XC_FLAGS_3D | XC_FLAGS_HAVE_EXC | XC_FLAGS_HAVE_VXC | XC_FLAGS_HAVE_FXC 
              | XC_FLAGS_HAVE_KXC, 

  ...
};
\end{verbatim}
Note that the references are separated by a newline. The user of the
library can access the information in the following way:
\begin{verbatim}
#include <stdio.h>
#include <xc.h>

int main()
{
  xc_func_type func;
  xc_func_init(&func, XC_GGA_X_B88, XC_UNPOLARIZED);
  printf("The functional '%s' is defined in the reference(s):\n%s\n",
    func.info->name, func.info->refs);
  xc_func_end(&func);
}
\end{verbatim}

\section{Conclusions and the future}

{\sc Libxc} is by now seven years old, and the code is quite stable
and in use by hundreds of scientists around the world. We are
committed to continue the development of the library in the future,
mainly by following the two lines mentioned before: include all
functionals, and their derivatives of the highest possible order. By
now essentially all LDA and GGA functionals ever proposed in the
literature are already included in the library. Unfortunately,
important gaps still remain especially in the meta-GGAs and hybrid
functionals. We will fill in those gaps in the near future. It is not
clear if {\sc Libxc} will ever include {\it all} functionals ever
developed (especially, as several new functionals come out every
year), but we will of course try\ldots

\section{Acknowledgements}

{\sc Libxc} profited considerably from other projects dedicated to xc
functionals (such as the density-functional repository in Daresbury of
H.\,J.\,J. van Dam), from generally available code (such as the
library of xc functionals of the Minnesota group --
\url{http://comp.chem.umn.edu/info/dft.htm}, xc routines of {\sc
  Abinit}~\cite{Gonze20092582,abinit2,Gonze2002478}, {\sc
  Espresso}~\cite{QE-2009}, etc.), and several individuals that
contributed either with code, bug fixes, or even bug reports. To all
of these we would like to express our gratitude.

MJTO thankfully acknowledges financial support from the Portuguese FCT 
(contract \#SFRH/BPD/44608/2008).

\appendix
\section{Available functionals}
\label{app:1}

For convenience, functionals in {\sc Libxc} are divided in exchange,
correlation, and exchange-correlation functionals. This division is
sometimes arbitrary, but it is often useful for the end user (who
may fancy strange mixtures of functionals) and for the code
developers.

Below we give a complete list of the functionals the library currently
knows about. We note that many of these were tested against reference
implementations, when available, or against published results. The
functionals have a label that is used to identify the functional
inside {\sc Libxc}. Furthermore, the third column indicates the time
in seconds required for 50,000,000 evaluations of the
(spin-unpolarized) xc potential in a single core of an Intel Core 2
processor running at 2.83 GHz. As one can see, the typical execution
time within {\sc Libxc} is much smaller than the total time required
in the whole calculation.

\begin{longtable}{lp{6cm}rl}
\caption{Functionals available in {\sc Libxc}.}\\

\multicolumn{3}{l}{\it LDA Functionals} \\
\\
\multicolumn{3}{l}{\hspace{1cm}\bf LDA Exchange} \\
XC\_LDA\_X & LDA exchange & 6.92 &\cite{CambridgeJournals:2040328,springerlink:10.1007/BF01340281} \\
XC\_LDA\_X\_2D & Slater exchange in 2D & 9.36\\
XC\_LDA\_X\_1D & Slater exchange in 1D &3382 & \cite{PhysRevA.83.032503} \\
\\
\multicolumn{3}{l}{\hspace{1cm}\bf LDA Correlation} \\
XC\_LDA\_C\_WIGNER & Wigner parametrization & 6.55 & \cite{TF9383400678} \\
XC\_LDA\_C\_RPA & Random Phase Approximation & 11.96 & \cite{PhysRev.106.364} \\
XC\_LDA\_C\_HL & Hedin \& Lundqvist & 9.86 & \cite{0022-3719-4-14-022} \\
XC\_LDA\_C\_GL & Gunnarsson \& Lundqvist & 12.19 & \cite{PhysRevB.13.4274} \\
XC\_LDA\_C\_XALPHA & Slater's X$\alpha$ (X-alpha) & 9.33\\
XC\_LDA\_C\_VWN & Vosko, Wilk, \& Nussair & 24.14 & \cite{vwn1980} \\
XC\_LDA\_C\_VWN\_RPA & Vosko, Wilk, \& Nussair (RPA) & 18.21 & \cite{vwn1980} \\
XC\_LDA\_C\_PZ & Perdew \& Zunger & 8.90 & \cite{PhysRevB.23.5048} \\
XC\_LDA\_C\_OB\_PZ & Ortiz \& Ballone (PZ parametrization) & 12.08 & \cite{PhysRevB.50.1391,PhysRevB.56.9970,PhysRevB.23.5048} \\
XC\_LDA\_C\_PW & Perdew \& Wang & 17.63 & \cite{PhysRevB.45.13244} \\
XC\_LDA\_C\_PW\_RPA & Perdew \& Wang fit to the RPA energy & 28.16 & \cite{PhysRevB.45.13244} \\
XC\_LDA\_C\_OB\_PW & Ortiz \& Ballone (PW parametrization) & 12.43 & \cite{PhysRevB.50.1391,PhysRevB.56.9970,PhysRevB.45.13244} \\
XC\_LDA\_C\_2D\_AMGB & Attaccalite, Moroni, Gori-Giorgi, and Bachelet (LDA for 2D systems) & 10.44 & \cite{PhysRevLett.88.256601} \\
XC\_LDA\_C\_2D\_PRM & Pittalis, R\"as\"anen, and Marques (LDA for 2D systems) & 11.35 & \cite{PhysRevB.78.195322} \\
XC\_LDA\_C\_vBH & von Barth \& Hedin & 10.40 & \cite{0022-3719-5-13-012} \\
XC\_LDA\_C\_1D\_CSC & Casula, Sorella \& Senatore (LDA correlation for 1D systems) & 28.85 & \cite{PhysRevB.74.245427} \\
XC\_LDA\_C\_ML1 & Modified LDA (version 1) of Proynov and Salahub & 37.02 & \cite{PhysRevB.49.7874} \\
XC\_LDA\_C\_ML2 & Modified LDA (version 2) of Proynov and Salahub & 51.99 & \cite{PhysRevB.49.7874} \\
XC\_LDA\_C\_GOMBAS & Gombas & 10.81 & \cite{gombas} \\
\\
\multicolumn{3}{l}{\hspace{1cm}\bf LDA Exchange-Correlation} \\
XC\_LDA\_XC\_TETER93 & Teter 1993 & 9.22 & \cite{PhysRevB.54.1703} \\
\\
\multicolumn{3}{l}{\hspace{1cm}\bf LDA Kinetic Energy} \\
XC\_LDA\_K\_TF & Thomas-Fermi kinetic energy & 6.74 & \cite{CambridgeJournals:1732980,Fermi1927} \\
XC\_LDA\_K\_LP & Lee and Parr Gaussian ansatz for the kinetic energy & 7.99 & \cite{PhysRevA.35.2377} \\
\\
\multicolumn{3}{l}{\it GGA Functionals} \\
\\
\multicolumn{3}{l}{\hspace{1cm}\bf GGA Exchange} \\
XC\_GGA\_X\_PBE & Perdew, Burke \& Ernzerhof exchange & 13.61 & \cite{PhysRevLett.77.3865,PhysRevLett.78.1396} \\
XC\_GGA\_X\_PBE\_R & Perdew, Burke \& Ernzerhof exchange (revised) & 19.27 & \cite{PhysRevLett.80.890} \\
XC\_GGA\_X\_MPBE & Adamo \& Barone modification to PBE & 19.39 & \cite{adamo:5933} \\
XC\_GGA\_X\_XPBE & Extended PBE by Xu \& Goddard III & 18.90 & \cite{xu:4068} \\
XC\_GGA\_X\_B86 & Becke 86 Xalfa,beta,gamma & 18.64 & \cite{becke:4524,becke:8554} \\
XC\_GGA\_X\_B86\_MGC & Becke 86 Xalfa,beta,gamma (with mod. grad. correction) & 24.86 & \cite{becke:4524,becke:7184} \\
XC\_GGA\_X\_B88 & Becke 88 & 20.97 & \cite{PhysRevA.38.3098} \\
XC\_GGA\_X\_G96 & Gill 96 & 15.57 & \cite{doi:10.1080/002689796173813} \\
XC\_GGA\_X\_PW86 & Perdew \& Wang 86 & 32.31 & \cite{PhysRevB.33.8800} \\
XC\_GGA\_X\_PW91 & Perdew \& Wang 91 & 48.02 & \cite{PhysRevB.46.6671} \\
XC\_GGA\_X\_OPTX & Handy \& Cohen OPTX 01 & 18.21 & \cite{doi:10.1080/00268970010018431} \\
XC\_GGA\_X\_DK87\_R1 & dePristo \& Kress 87 (version R1) & 19.10 & \cite{depristo:1425} \\
XC\_GGA\_X\_DK87\_R2 & dePristo \& Kress 87 (version R2) & 24.11 & \cite{depristo:1425} \\
XC\_GGA\_X\_LG93 & Lacks \& Gordon 93 & 32.74 & \cite{PhysRevA.47.4681} \\
XC\_GGA\_X\_FT97\_A & Filatov \& Thiel 97 (version A) & 22.05 & \cite{doi:10.1080/002689797170950} \\
XC\_GGA\_X\_FT97\_B & Filatov \& Thiel 97 (version B) & 26.86 & \cite{doi:10.1080/002689797170950} \\
XC\_GGA\_X\_PBE\_SOL & Perdew, Burke \& Ernzerhof exchange (for solids) & 19.33 & \cite{PhysRevLett.100.136406} \\
XC\_GGA\_X\_RPBE & Hammer, Hansen \& Norskov (PBE-like) & 17.14 & \cite{PhysRevB.59.7413} \\
XC\_GGA\_X\_WC & Wu \& Cohen & 17.91 & \cite{PhysRevB.73.235116} \\
XC\_GGA\_X\_AM05 & Armiento \& Mattsson 05 exchange & 61.38 & \cite{PhysRevB.72.085108,mattsson:084714} \\
XC\_GGA\_X\_PBEA & Madsen 07 & 30.98 & \cite{PhysRevB.75.195108} \\
XC\_GGA\_X\_mPW91 & mPW91 of Adamo \& Barone & 47.19 & \cite{adamo:664} \\
XC\_GGA\_X\_2D\_B86\_MGC & Becke 86 with modified gradient correction for 2D & 13.49 & \cite{PhysRevA.79.012503} \\
XC\_GGA\_X\_BAYESIAN & Bayesian best fit for the enhancement factor & 18.91 & \cite{PhysRevLett.95.216401} \\
XC\_GGA\_X\_PBE\_JSJR & Reparametrized PBE by Pedroza, Silva \& Capelle & 18.71 &\cite{PhysRevB.79.201106} \\
XC\_GGA\_X\_OPTB88\_VDW & opt-Becke 88 for vdW & 25.24 & \cite{0953-8984-22-2-022201} \\
XC\_GGA\_X\_PBEK1\_VDW & Reparametrized PBE for vdW & 18.72 & \cite{0953-8984-22-2-022201} \\
XC\_GGA\_X\_OPTPBE\_VDW & Reparametrized PBE for vdW & 40.62 & \cite{0953-8984-22-2-022201} \\
XC\_GGA\_X\_RGE2 & Regularized PBE & 14.65 & \cite{doi:10.1021/ct8005369} \\
XC\_GGA\_X\_RPW86 & Refitted Perdew \& Wang 86 & 30.61 & \cite{doi:10.1021/ct900365q} \\
XC\_GGA\_X\_KT1 & Keal and Tozer, version 1 & 23.96 & \cite{keal:3015} \\
XC\_GGA\_X\_HERMAN & Herman Xalphabeta GGA & 17.44 & \cite{PhysRevLett.22.807,QUA:QUA560040746} \\
XC\_GGA\_X\_LBM & van Leeuwen \& Baerends modified & 30.98 & \cite{schipper:1344}  \\
XC\_GGA\_X\_OL2 & Exchange form based on Ou-Yang and Levy v.2 & 15.69 & \cite{Fuentealba199531,QUA:QUA560400309} \\
XC\_GGA\_X\_MB88 & Modified Becke 88 for proton transfer & 24.86 & \cite{doi:10.1021/jp903672e} \\
XC\_GGA\_X\_APBE & mu fixed from the semiclassical neutral atom & 18.71 & \cite{PhysRevLett.106.186406} \\
XC\_GGA\_X\_HTBS & Haas, Tran, Blaha, and Schwarz & 26.17 & \cite{PhysRevB.83.205117} \\
XC\_GGA\_X\_AIRY & Constantin et al based on the Airy gas & 47.48 & \cite{PhysRevB.80.035125} \\
XC\_GGA\_X\_LAG & Local Airy Gas & 49.75 & \cite{PhysRevB.62.10046} \\
XC\_GGA\_X\_SOGGA11 & Second-order generalized gradient approximation 2011 & 20.40 & \cite{doi:10.1021/jz200616w} \\
XC\_GGA\_X\_C09X & C09x to be used with the VdW of Rutgers-Chalmers & 19.86 & \cite{PhysRevB.81.161104} \\
\\
\multicolumn{3}{l}{\hspace{1cm}\bf GGA Correlation} \\
XC\_GGA\_C\_PBE & Perdew, Burke \& Ernzerhof correlation & 39.69 & \cite{PhysRevLett.77.3865,PhysRevLett.78.1396} \\
XC\_GGA\_C\_XPBE & Extended PBE by Xu \& Goddard III & 43.85 & \cite{xu:4068} \\
XC\_GGA\_C\_LYP & Lee, Yang \& Parr & 25.88 & \cite{PhysRevB.37.785,Miehlich1989200} \\
XC\_GGA\_C\_P86 & Perdew 86 & 51.36 & \cite{PhysRevB.33.8822} \\
XC\_GGA\_C\_PBE\_SOL & Perdew, Burke \& Ernzerhof correlation SOL & 55.32 & \cite{PhysRevLett.100.136406} \\
XC\_GGA\_C\_PW91 & Perdew \& Wang 91 & 51.17 & \cite{PhysRevB.46.6671,PhysRevB.48.4978.2} \\
XC\_GGA\_C\_AM05 & Armiento \& Mattsson 05 correlation & 25.88 & \cite{PhysRevB.72.085108} \\
XC\_GGA\_C\_LM & Langreth \& Mehl & 42.87 & \cite{PhysRevLett.47.446} \\
XC\_GGA\_C\_PBE\_JRGX & Reparametrized PBE by Pedroza, Silva \& Capelle & 55.74 & \cite{PhysRevB.79.201106} \\
XC\_GGA\_C\_RGE2 & Regularized PBE & 40.55 & \cite{doi:10.1021/ct8005369} \\
XC\_GGA\_C\_WL & Wilson \& Levy & 20.95 & \cite{PhysRevB.41.12930} \\
XC\_GGA\_C\_WI & Wilson \& Ivanov & 16.48 & \cite{QUA:QUA9} \\
XC\_GGA\_C\_WI0 & Wilson \& Ivanov initial version & 19.42 & \cite{QUA:QUA9} \\
XC\_GGA\_C\_APBE & mu fixed from the semiclassical neutral atom & 40.86 & \cite{PhysRevLett.106.186406} \\
XC\_GGA\_C\_SOGGA11 & Second-order generalized gradient approximation 2011 & 40.67 & \cite{doi:10.1021/jz200616w} \\
XC\_GGA\_C\_SOGGA11\_X & to be used with XC\_HYB\_GGA\_X\_SOGGA11\_X & 37.05 & \cite{peverati:191102} \\
\\
\multicolumn{3}{l}{\hspace{1cm}\bf GGA Exchange-Correlation} \\
XC\_GGA\_XC\_LB & van Leeuwen \& Baerends & 29.27 & \cite{PhysRevA.49.2421} \\
XC\_GGA\_XC\_HCTH\_93 & HCTH functional fitted to  93 molecules & 155.10 & \cite{hamprecht:6264} \\
XC\_GGA\_XC\_HCTH\_120 & HCTH functional fitted to 120 molecules & 151.57 & \cite{boese:1670} \\
XC\_GGA\_XC\_HCTH\_147 & HCTH functional fitted to 147 molecules & 143.57 & \cite{boese:1670} \\
XC\_GGA\_XC\_HCTH\_407 & HCTH functional fitted to 147 molecules & 155.60 & \cite{boese:5497} \\
XC\_GGA\_XC\_EDF1 & Empirical functional from Adamson, Gill, and Pople & 69.03 & \cite{Adamson19986} \\
XC\_GGA\_XC\_XLYP & XLYP functional & 106.30 & \cite{Xu02032004} \\
XC\_GGA\_XC\_PBE1W & PBE1W (functional fitted for water) & 94.72 & \cite{doi:10.1021/jp052436c} \\
XC\_GGA\_XC\_MPWLYP1W & mPWLYP1w (functional fitted for water) & 106.74 & \cite{doi:10.1021/jp052436c} \\
XC\_GGA\_XC\_PBELYP1W & PBELYP1W (functional fitted for water) & 79.12 & \cite{doi:10.1021/jp052436c} \\
XC\_GGA\_XC\_KT2 & Keal and Tozer, version 2 & 31.73 & \cite{keal:3015} \\
XC\_GGA\_XC\_TH\_FL & Tozer and Handy v. FL & 94.64 & \cite{Tozer1997183} \\
XC\_GGA\_XC\_TH\_FC & Tozer and Handy v. FC & 247.35 & \cite{Tozer1997183} \\
XC\_GGA\_XC\_TH\_FCFO & Tozer and Handy v. FCFO & 372.59 & \cite{Tozer1997183} \\
XC\_GGA\_XC\_TH\_FCO & Tozer and Handy v. FCO & 355.16 & \cite{Tozer1997183} \\
XC\_GGA\_XC\_TH1 & Tozer and Handy v. 1 & 383.44 & \cite{tozer:2545} \\
XC\_GGA\_XC\_TH2 & Tozer and Handy v. 2 & 352.52 & \cite{doi:10.1021/jp980259s} \\
XC\_GGA\_XC\_TH3 & Tozer and Handy v. 3 & 332.24 & \cite{doi:10.1080/002689798167863} \\
XC\_GGA\_XC\_TH3 & Tozer and Handy v. 4 & 337.86 & \cite{doi:10.1080/002689798167863} \\
\\
\multicolumn{3}{l}{\hspace{1cm}\bf GGA Kinetic Energy} \\
XC\_GGA\_K\_VW & von Weiszaecker correction to Thomas-Fermi & 18.49 & \cite{springerlink:10.1007/BF01337700} \\
XC\_GGA\_K\_GE2 & Second-order gradient expansion of the kinetic energy density & 16.08 & \cite{Kompaneets1956,Kirznits1957} \\
XC\_GGA\_K\_GOLDEN & TF-lambda-vW form by Golden ($l = 13/45$) & 16.39 & \cite{PhysRev.105.604} \\
XC\_GGA\_K\_YT65 & TF-lambda-vW form by Yonei and Tomishima ($l = 1/5$) & 13.29 & \cite{JPSJ.20.1051} \\
XC\_GGA\_K\_BALTIN & TF-lambda-vW form by Baltin ($l = 5/9$) & 13.24 & \cite{Baltin1972} \\
XC\_GGA\_K\_LIEB & TF-lambda-vW form by Lieb ($l = 0.185909191$) & 14.68 & \cite{RevModPhys.53.603} \\
XC\_GGA\_K\_ABSR1 & gamma-TFvW form by Acharya et al [$g = 1 - 1.412/N^{1/3}$] & 17.54 & \cite{Acharya01121980} \\
XC\_GGA\_K\_ABSR2 & gamma-TFvW form by Acharya et al [$g = 1 - 1.332/N^{1/3}$] & 17.01 & \cite{Acharya01121980} \\
XC\_GGA\_K\_GR & gamma-TFvW form by G\'azquez and Robles & 13.89 & \cite{gazquez:1467} \\
XC\_GGA\_K\_LUDENA & gamma-TFvW form by Lude\~na & 18.72 & \cite{Ludena1986} \\
XC\_GGA\_K\_GP85 & gamma-TFvW form by Ghosh and Parr & 18.19 & \cite{ghosh:3307} \\
XC\_GGA\_K\_LLP & Becke 88 & 22.69 & \cite{PhysRevA.44.768} \\
XC\_GGA\_K\_FR\_B88 & Fuentealba \& Reyes (B88 version) & 19.97 & \cite{Fuentealba199531} \\
XC\_GGA\_K\_FR\_PW86 & Fuentealba \& Reyes (PW86 version) & 32.24 & \cite{Fuentealba199531} \\
XC\_GGA\_K\_PEARSON & Pearson 1992 & 13.30 & \cite{lacks:4446,pearson:881} \\
XC\_GGA\_K\_OL1 & Ou-Yang and Levy v.1 & 20.43 & \cite{QUA:QUA560400309} \\
XC\_GGA\_K\_OL2 & Ou-Yang and Levy v.2 & 20.26 & \cite{QUA:QUA560400309} \\
XC\_GGA\_K\_DK & DePristo and Kress & 18.64 & \cite{PhysRevA.35.438} \\
XC\_GGA\_K\_PERDEW & Perdew & 16.21 & \cite{JohnP199279} \\
XC\_GGA\_K\_VSK & Vitos, Skriver, and Kollar & 17.05 & \cite{PhysRevB.57.12611} \\
XC\_GGA\_K\_VJKS & Vitos, Johansson, Kollar, and Skriver & 18.61 & \cite{PhysRevA.61.052511} \\
XC\_GGA\_K\_ERNZERHOF & Ernzerhof & 18.72 & \cite{M200059} \\
XC\_GGA\_K\_LC94 & Lembarki \& Chermette & 46.63 & \cite{PhysRevA.50.5328} \\
XC\_GGA\_K\_APBE & mu fixed from the semiclassical neutral atom & 19.48 & \cite{PhysRevLett.106.186406} \\
XC\_GGA\_K\_THAKKAR & Thakkar 1992 & 24.92 & \cite{PhysRevA.46.6920} \\
XC\_GGA\_K\_TW1 & Tran and Wesolowski set 1 (Table II) & 18.70 & \cite{QUA:QUA10306} \\
XC\_GGA\_K\_TW2 & Tran and Wesolowski set 2 (Table II) & 18.30 & \cite{QUA:QUA10306} \\
XC\_GGA\_K\_TW3 & Tran and Wesolowski set 3 (Table II) & 19.01 & \cite{QUA:QUA10306} \\
XC\_GGA\_K\_TW4 & Tran and Wesolowski set 4 (Table II) & 19.97 & \cite{QUA:QUA10306} \\
\\
\multicolumn{3}{l}{\it MetaGGA Functionals} \\
\\
\multicolumn{3}{l}{\hspace{1cm}\bf MetaGGA Exchange} \\
XC\_MGGA\_X\_LTA & Local tau approximation & 24.07 & \cite{ernzerhof:911} \\
XC\_MGGA\_X\_TPSS & Perdew, Tao, Staroverov \& Scuseria exchange & 28.88 & \cite{PhysRevLett.91.146401,perdew:6898} \\
XC\_MGGA\_X\_TAU\_HCTH, & Tau HCTH & 23.45 & \cite{boese:9559} \\
XC\_MGGA\_X\_GVT4 & GVT4 (exchange part of VSXC) & 26.85 & \cite{voorhis:400} \\
XC\_MGGA\_X\_M06L & M06-L & 29.74 & \cite{zhao:194101,springerlink:10.1007/s00214-007-0401-8} \\
XC\_MGGA\_X\_BR89 & Becke-Roussel 89 & 49.39 & \cite{PhysRevA.39.3761} \\
XC\_MGGA\_X\_BJ06 & Becke \& Johnson 06 & 35.28 & \cite{becke:221101} \\
XC\_MGGA\_X\_TB09 & Tran \& Blaha 09 & 31.96 & \cite{PhysRevLett.102.226401} \\
XC\_MGGA\_X\_RPP09 & Rasanen, Pittalis \& Proetto 09 & 44.16 & \cite{rasanen:044112} \\
XC\_MGGA\_X\_2D\_PRHG07 & Pittalis-Rasanen-Helbig-Gross 2007 & 54.18 & \cite{PhysRevB.76.235314} \\
XC\_MGGA\_X\_2D\_PRHG07\_PRP10 & Corrected Pittalis-Rasanen-Helbig-Gross 2010 & 53.94 & \cite{PhysRevB.76.235314} \\
\\
\multicolumn{3}{l}{\hspace{1cm}\bf MetaGGA Correlation} \\
XC\_MGGA\_C\_TPSS & Perdew, Tao, Staroverov \& Scuseria correlation & 184.60 & \cite{PhysRevLett.91.146401,perdew:6898} \\
XC\_MGGA\_C\_VSXC & VSXC (correlation part) & 148.84 & \cite{voorhis:400}
\\
\multicolumn{3}{l}{\it Hybrid Functionals} \\
\\
\multicolumn{3}{l}{\hspace{1cm}\bf Hybrid Exchange} \\
XC\_HYB\_GGA\_X\_SOGGA11\_X & Hybrid based on SOGGA11 form (see also XC\_GGA\_C\_SOGGA11\_X) & 21.13 & \cite{peverati:191102} \\
\\
\multicolumn{3}{l}{\hspace{1cm}\bf Hybrid Exchange-Correlation} \\
XC\_HYB\_GGA\_XC\_B3PW91  & The original hybrid proposed by Becke & 99.90& \cite{becke:5648} \\
XC\_HYB\_GGA\_XC\_B3LYP & The (in)famous B3LYP & 93.55 & \cite{doi:10.1021/j100096a001} \\
XC\_HYB\_GGA\_XC\_B3P86 & Perdew 86 hybrid similar to B3PW91 & 105.17 \\
XC\_HYB\_GGA\_XC\_O3LYP & hybrid using the optx functional & 78.60 & \cite{doi:10.1080/00268970010023435} \\ 
XC\_HYB\_GGA\_XC\_PBEH & PBEH, also known as PBE0 & 76.12 & \cite{ernzerhof:5029} \\
XC\_HYB\_GGA\_XC\_X3LYP & X3LYP & 120.61 & \cite{Xu02032004} \\
XC\_HYB\_GGA\_XC\_B1WC & B1WC & 78.94 & \cite{PhysRevB.77.165107} \\
XC\_HYB\_GGA\_XC\_B97 & Becke 97 & 158.49 & \cite{becke:8554} \\
XC\_HYB\_GGA\_XC\_B97\_1 & Becke 97-1 & 154.97 & \cite{hamprecht:6264} \\
XC\_HYB\_GGA\_XC\_B97\_2 & Becke 97-2 & 151.12 & \cite{wilson:9233} \\
XC\_HYB\_GGA\_XC\_B97\_K & Becke 97-K, Boese-Martin for Kinetics & 158.38 & \cite{boese:3405} \\
XC\_HYB\_GGA\_XC\_B97\_3 & Becke 97-3 & 155.92 & \cite{keal:121103} \\
XC\_HYB\_GGA\_XC\_B1LYP & B1LYP & 52.18 & \cite{Adamo1997242} \\
XC\_HYB\_GGA\_XC\_B1PW91 & B1PW91 & 69.92 & \cite{Adamo1997242} \\
XC\_HYB\_GGA\_XC\_mPW1PW & mPW1PW & 99.96 & \cite{adamo:664} \\
XC\_HYB\_GGA\_XC\_mPW3PW & mPW3PW of Adamo \& Barone & 110.82 & \cite{adamo:664} \\
XC\_HYB\_GGA\_XC\_mPW3LYP & mPW3LYP & 112.92 & \cite{doi:10.1021/jp048147q} \\
XC\_HYB\_GGA\_XC\_SB98\_1a & SB98 (1a) & 149.81 & \cite{schmider:9624} \\
XC\_HYB\_GGA\_XC\_SB98\_1b & SB98 (1b) & 156.87 & \cite{schmider:9624} \\
XC\_HYB\_GGA\_XC\_SB98\_1c & SB98 (1c) & 158.72 & \cite{schmider:9624} \\
XC\_HYB\_GGA\_XC\_SB98\_2a & SB98 (2a) & 139.63 & \cite{schmider:9624} \\
XC\_HYB\_GGA\_XC\_SB98\_2b & SB98 (2b) & 148.48 & \cite{schmider:9624} \\
XC\_HYB\_GGA\_XC\_SB98\_2c & SB98 (2c) & 158.78 & \cite{schmider:9624} \\
XC\_HYB\_GGA\_XC\_mPW1K & mPW1K & 100.53 & \cite{doi:10.1021/jp000497z} \\
\end{longtable}


\bibliographystyle{model1a-num-names}

\begin{thebibliography}{164}
\expandafter\ifx\csname natexlab\endcsname\relax\def\natexlab#1{#1}\fi
\providecommand{\bibinfo}[2]{#2}
\ifx\xfnm\relax \def\xfnm[#1]{\unskip,\space#1}\fi
\bibitem[{Hohenberg and Kohn(1964)}]{PhysRev.136.B864}
\bibinfo{author}{P.~Hohenberg}, \bibinfo{author}{W.~Kohn},
  \bibinfo{journal}{Phys. Rev.} \bibinfo{volume}{136} (\bibinfo{year}{1964})
  \bibinfo{pages}{B864}.
\bibitem[{Parr and Yang(1989)}]{Parr1989}
\bibinfo{author}{R.~G. Parr}, \bibinfo{author}{W.~Yang},
  \bibinfo{title}{Density-Functional Theory of Atoms and Molecules},
  \bibinfo{publisher}{Oxford Univ. Press}, \bibinfo{address}{New York},
  \bibinfo{year}{1989}.
\bibitem[{Dreizler and Gross(1990)}]{Dreizler90}
\bibinfo{author}{R.~Dreizler}, \bibinfo{author}{E.~Gross},
  \bibinfo{title}{{D}ensity {F}unctional {T}heory},
  \bibinfo{publisher}{Springer}, \bibinfo{address}{Berlin},
  \bibinfo{year}{1990}.
\bibitem[{Marques et~al.(2003)Marques, Fiolhais, and Marques}]{Fiolhais2003}
\bibinfo{editor}{M.~Marques}, \bibinfo{editor}{C.~Fiolhais},
  \bibinfo{editor}{M.~Marques} (Eds.), \bibinfo{title}{A Primer in Density
  Functional Theory}, \bibinfo{publisher}{Springer}, \bibinfo{address}{Berlin},
  \bibinfo{year}{2003}.
\bibitem[{Kohn and Sham(1965)}]{PhysRev.140.A1133}
\bibinfo{author}{W.~Kohn}, \bibinfo{author}{L.~J. Sham},
  \bibinfo{journal}{Phys. Rev.} \bibinfo{volume}{140} (\bibinfo{year}{1965})
  \bibinfo{pages}{A1133}.
\bibitem[{van Leeuwen(2003)}]{vanLeeuwen200325}
\bibinfo{author}{R.~van Leeuwen}, volume~\bibinfo{volume}{43} of
  \textit{\bibinfo{series}{Advances in Quantum Chemistry}},
  \bibinfo{publisher}{Academic Press}, \bibinfo{year}{2003}, pp.
  \bibinfo{pages}{25 -- 94}.
\bibitem[{Ceperley and Alder(1980)}]{PhysRevLett.45.566}
\bibinfo{author}{D.~M. Ceperley}, \bibinfo{author}{B.~J. Alder},
  \bibinfo{journal}{Phys. Rev. Lett.} \bibinfo{volume}{45}
  (\bibinfo{year}{1980}) \bibinfo{pages}{566}.
\bibitem[{Ortiz and Ballone(1994)}]{PhysRevB.50.1391}
\bibinfo{author}{G.~Ortiz}, \bibinfo{author}{P.~Ballone},
  \bibinfo{journal}{Phys. Rev. B} \bibinfo{volume}{50} (\bibinfo{year}{1994})
  \bibinfo{pages}{1391}.
\bibitem[{Scuseria and Staroverov(2005)}]{Scuseria2005}
\bibinfo{author}{G.~E. Scuseria}, \bibinfo{author}{V.~N. Staroverov}, in:
  \bibinfo{editor}{C.~E. Dykstra}, \bibinfo{editor}{G.~Frenking},
  \bibinfo{editor}{K.~S. Kim}, \bibinfo{editor}{G.~E. Scuseria} (Eds.),
  \bibinfo{booktitle}{Theory and Applications of Computational Chemistry: The
  First Forty Years}, \bibinfo{publisher}{Elsevier},
  \bibinfo{address}{Amsterdam}, \bibinfo{year}{2005}, p. \bibinfo{pages}{669}.
\bibitem[{Perdew and Schmidt(2001)}]{perdew:1}
\bibinfo{author}{J.~P. Perdew}, \bibinfo{author}{K.~Schmidt},
  \bibinfo{journal}{AIP Conf. Proc.} \bibinfo{volume}{577}
  (\bibinfo{year}{2001}) \bibinfo{pages}{1}.
\bibitem[{Thomas(1927)}]{CambridgeJournals:1732980}
\bibinfo{author}{L.~H. Thomas}, \bibinfo{journal}{Math. Proc. Cambridge Philos.
  Soc.} \bibinfo{volume}{23} (\bibinfo{year}{1927}) \bibinfo{pages}{542}.
\bibitem[{Fermi(1927)}]{Fermi1927}
\bibinfo{author}{E.~Fermi}, \bibinfo{journal}{Rend. Accad. Naz. Lincei}
  \bibinfo{volume}{6} (\bibinfo{year}{1927}) \bibinfo{pages}{602}.
\bibitem[{Chen and Zhou(2008)}]{Chen2008}
\bibinfo{author}{H.~Chen}, \bibinfo{author}{A.~Zhou}, \bibinfo{journal}{Numer.
  Math. Theor. Meth. Appl.} \bibinfo{volume}{1} (\bibinfo{year}{2008})
  \bibinfo{pages}{1}.
\bibitem[{Lude\~na and Karasiev(202)}]{Ludena2002}
\bibinfo{author}{E.~V. Lude\~na}, \bibinfo{author}{V.~V. Karasiev}, in:
  \bibinfo{editor}{K.~D. Sen} (Ed.), \bibinfo{booktitle}{Reviews of Modern
  Quantum Chemistry: A Celebration of the Contributions of Robert Parr},
  volume~\bibinfo{volume}{1}, \bibinfo{publisher}{World Scientific},
  \bibinfo{address}{Singapore}, \bibinfo{year}{202}, p. \bibinfo{pages}{612}.
\bibitem[{Marques et~al.(2003)Marques, Castro, Bertsch, and
  Rubio}]{Marques200360}
\bibinfo{author}{M.~A. Marques}, \bibinfo{author}{A.~Castro},
  \bibinfo{author}{G.~F. Bertsch}, \bibinfo{author}{A.~Rubio},
  \bibinfo{journal}{Comp. Phys. Comm.} \bibinfo{volume}{151}
  (\bibinfo{year}{2003}) \bibinfo{pages}{60}.
\bibitem[{Castro et~al.(2006)Castro, Appel, Oliveira, Rozzi, Andrade, Lorenzen,
  Marques, Gross, and Rubio}]{PSSB:PSSB200642067}
\bibinfo{author}{A.~Castro}, \bibinfo{author}{H.~Appel},
  \bibinfo{author}{M.~Oliveira}, \bibinfo{author}{C.~A. Rozzi},
  \bibinfo{author}{X.~Andrade}, \bibinfo{author}{F.~Lorenzen},
  \bibinfo{author}{M.~A.~L. Marques}, \bibinfo{author}{E.~K.~U. Gross},
  \bibinfo{author}{A.~Rubio}, \bibinfo{journal}{Phys. Status Solidi B}
  \bibinfo{volume}{243} (\bibinfo{year}{2006}) \bibinfo{pages}{2465}.
\bibitem[{Andrade et~al.(2007)Andrade, Botti, Marques, and
  Rubio}]{andrade:184106}
\bibinfo{author}{X.~Andrade}, \bibinfo{author}{S.~Botti},
  \bibinfo{author}{M.~A.~L. Marques}, \bibinfo{author}{A.~Rubio},
  \bibinfo{journal}{J. Chem. Phys.} \bibinfo{volume}{126}
  (\bibinfo{year}{2007}) \bibinfo{pages}{184106}.
\bibitem[{Castro et~al.(2004)Castro, Marques, Alonso, and
  Rubio}]{Alberto:2004:1546-1955:231}
\bibinfo{author}{A.~Castro}, \bibinfo{author}{M.~A. Marques},
  \bibinfo{author}{J.~A. Alonso}, \bibinfo{author}{A.~Rubio},
  \bibinfo{journal}{J. Comp. Theo. Nanosci.} \bibinfo{volume}{1}
  (\bibinfo{year}{2004}) \bibinfo{pages}{231}.
\bibitem[{Strange et~al.(2001)Strange, Manby, and Knowles}]{Strange2001310}
\bibinfo{author}{R.~Strange}, \bibinfo{author}{F.~Manby},
  \bibinfo{author}{P.~Knowles}, \bibinfo{journal}{Comp. Phys. Comm.}
  \bibinfo{volume}{136} (\bibinfo{year}{2001}) \bibinfo{pages}{310}.
\bibitem[{Sałek and Hesselmann(2007)}]{JCC:JCC20758}
\bibinfo{author}{P.~Sałek}, \bibinfo{author}{A.~Hesselmann},
  \bibinfo{journal}{J. Comp. Chem.} \bibinfo{volume}{28} (\bibinfo{year}{2007})
  \bibinfo{pages}{2569}.
\bibitem[{Gonze et~al.(2009)Gonze, Amadon, Anglade, Beuken, Bottin, Boulanger,
  Bruneval, Caliste, Caracas, Côté, Deutsch, Genovese, Ghosez, Giantomassi,
  Goedecker, Hamann, Hermet, Jollet, Jomard, Leroux, Mancini, Mazevet,
  Oliveira, Onida, Pouillon, Rangel, Rignanese, Sangalli, Shaltaf, Torrent,
  Verstraete, Zerah, and Zwanziger}]{Gonze20092582}
\bibinfo{author}{X.~Gonze}, \bibinfo{author}{B.~Amadon}, \bibinfo{author}{P.-M.
  Anglade}, \bibinfo{author}{J.-M. Beuken}, \bibinfo{author}{F.~Bottin},
  \bibinfo{author}{P.~Boulanger}, \bibinfo{author}{F.~Bruneval},
  \bibinfo{author}{D.~Caliste}, \bibinfo{author}{R.~Caracas},
  \bibinfo{author}{M.~Côté}, \bibinfo{author}{T.~Deutsch},
  \bibinfo{author}{L.~Genovese}, \bibinfo{author}{P.~Ghosez},
  \bibinfo{author}{M.~Giantomassi}, \bibinfo{author}{S.~Goedecker},
  \bibinfo{author}{D.~Hamann}, \bibinfo{author}{P.~Hermet},
  \bibinfo{author}{F.~Jollet}, \bibinfo{author}{G.~Jomard},
  \bibinfo{author}{S.~Leroux}, \bibinfo{author}{M.~Mancini},
  \bibinfo{author}{S.~Mazevet}, \bibinfo{author}{M.~Oliveira},
  \bibinfo{author}{G.~Onida}, \bibinfo{author}{Y.~Pouillon},
  \bibinfo{author}{T.~Rangel}, \bibinfo{author}{G.-M. Rignanese},
  \bibinfo{author}{D.~Sangalli}, \bibinfo{author}{R.~Shaltaf},
  \bibinfo{author}{M.~Torrent}, \bibinfo{author}{M.~Verstraete},
  \bibinfo{author}{G.~Zerah}, \bibinfo{author}{J.~Zwanziger},
  \bibinfo{journal}{Comp. Phys. Comm.} \bibinfo{volume}{180}
  (\bibinfo{year}{2009}) \bibinfo{pages}{2582}.
\bibitem[{Gonze et~al.(2005)Gonze, Rignanese, Verstraete, Beuken, Pouillon,
  Caracas, Jollet, Torrent, Zerah, Mikami, Ghosez, Veithen, Raty, Olevano,
  Bruneval, Reining, Godby, Onida, Hamann, and Allan}]{abinit2}
\bibinfo{author}{X.~Gonze}, \bibinfo{author}{G.-M. Rignanese},
  \bibinfo{author}{M.~Verstraete}, \bibinfo{author}{J.-M. Beuken},
  \bibinfo{author}{Y.~Pouillon}, \bibinfo{author}{R.~Caracas},
  \bibinfo{author}{F.~Jollet}, \bibinfo{author}{M.~Torrent},
  \bibinfo{author}{G.~Zerah}, \bibinfo{author}{M.~Mikami},
  \bibinfo{author}{P.~Ghosez}, \bibinfo{author}{M.~Veithen},
  \bibinfo{author}{J.-Y. Raty}, \bibinfo{author}{V.~Olevano},
  \bibinfo{author}{F.~Bruneval}, \bibinfo{author}{L.~Reining},
  \bibinfo{author}{R.~Godby}, \bibinfo{author}{G.~Onida},
  \bibinfo{author}{D.~Hamann}, \bibinfo{author}{D.~Allan},
  \bibinfo{journal}{Zeit. Kristallogr.} \bibinfo{volume}{220}
  (\bibinfo{year}{2005}) \bibinfo{pages}{558}.
\bibitem[{Gonze et~al.(2002)Gonze, Beuken, Caracas, Detraux, Fuchs, Rignanese,
  Sindic, Verstraete, Zerah, Jollet, Torrent, Roy, Mikami, Ghosez, Raty, and
  Allan}]{Gonze2002478}
\bibinfo{author}{X.~Gonze}, \bibinfo{author}{J.-M. Beuken},
  \bibinfo{author}{R.~Caracas}, \bibinfo{author}{F.~Detraux},
  \bibinfo{author}{M.~Fuchs}, \bibinfo{author}{G.-M. Rignanese},
  \bibinfo{author}{L.~Sindic}, \bibinfo{author}{M.~Verstraete},
  \bibinfo{author}{G.~Zerah}, \bibinfo{author}{F.~Jollet},
  \bibinfo{author}{M.~Torrent}, \bibinfo{author}{A.~Roy},
  \bibinfo{author}{M.~Mikami}, \bibinfo{author}{P.~Ghosez},
  \bibinfo{author}{J.-Y. Raty}, \bibinfo{author}{D.~Allan},
  \bibinfo{journal}{Comp. Mat. Sci.} \bibinfo{volume}{25}
  (\bibinfo{year}{2002}) \bibinfo{pages}{478}.
\bibitem[{Oliveira and Nogueira(2008)}]{Oliveira2008524}
\bibinfo{author}{M.~J. Oliveira}, \bibinfo{author}{F.~Nogueira},
  \bibinfo{journal}{Comp. Phys. Comm.} \bibinfo{volume}{178}
  (\bibinfo{year}{2008}) \bibinfo{pages}{524}.
\bibitem[{Holzwarth et~al.(2001)Holzwarth, Tackett, and
  Matthews}]{Holzwarth2001329}
\bibinfo{author}{N.~Holzwarth}, \bibinfo{author}{A.~Tackett},
  \bibinfo{author}{G.~Matthews}, \bibinfo{journal}{Comp. Phys. Comm.}
  \bibinfo{volume}{135} (\bibinfo{year}{2001}) \bibinfo{pages}{329}.
\bibitem[{Brandbyge et~al.(2002)Brandbyge, Mozos, Ordej\'on, Taylor, and
  Stokbro}]{PhysRevB.65.165401}
\bibinfo{author}{M.~Brandbyge}, \bibinfo{author}{J.-L. Mozos},
  \bibinfo{author}{P.~Ordej\'on}, \bibinfo{author}{J.~Taylor},
  \bibinfo{author}{K.~Stokbro}, \bibinfo{journal}{Phys. Rev. B}
  \bibinfo{volume}{65} (\bibinfo{year}{2002}) \bibinfo{pages}{165401}.
\bibitem[{Soler et~al.(2002)Soler, Artacho, Gale, Garc\'{\i}a, Junquera,
  Ordej\'on, and S\'anchez-Portal}]{0953-8984-14-11-302}
\bibinfo{author}{J.~M. Soler}, \bibinfo{author}{E.~Artacho},
  \bibinfo{author}{J.~D. Gale}, \bibinfo{author}{A.~Garc\'{\i}a},
  \bibinfo{author}{J.~Junquera}, \bibinfo{author}{P.~Ordej\'on},
  \bibinfo{author}{D.~S\'anchez-Portal}, \bibinfo{journal}{J. Phys.: Cond.
  Matt.} \bibinfo{volume}{14} (\bibinfo{year}{2002}) \bibinfo{pages}{2745}.
\bibitem[{Genovese et~al.(2011)Genovese, Videau, Ospici, Deutsch, Goedecker,
  and Méhaut}]{Genovese2011149}
\bibinfo{author}{L.~Genovese}, \bibinfo{author}{B.~Videau},
  \bibinfo{author}{M.~Ospici}, \bibinfo{author}{T.~Deutsch},
  \bibinfo{author}{S.~Goedecker}, \bibinfo{author}{J.-F. Méhaut},
  \bibinfo{journal}{Comptes Rendus M\'ecanique} \bibinfo{volume}{339}
  (\bibinfo{year}{2011}) \bibinfo{pages}{149}.
\bibitem[{Genovese et~al.(2008)Genovese, Neelov, Goedecker, Deutsch, Ghasemi,
  Willand, Caliste, Zilberberg, Rayson, Bergman, and
  Schneider}]{genovese:014109}
\bibinfo{author}{L.~Genovese}, \bibinfo{author}{A.~Neelov},
  \bibinfo{author}{S.~Goedecker}, \bibinfo{author}{T.~Deutsch},
  \bibinfo{author}{S.~A. Ghasemi}, \bibinfo{author}{A.~Willand},
  \bibinfo{author}{D.~Caliste}, \bibinfo{author}{O.~Zilberberg},
  \bibinfo{author}{M.~Rayson}, \bibinfo{author}{A.~Bergman},
  \bibinfo{author}{R.~Schneider}, \bibinfo{journal}{J. Chem. Phys.}
  \bibinfo{volume}{129} (\bibinfo{year}{2008}) \bibinfo{pages}{014109}.
\bibitem[{Sottile et~al.(2005)Sottile, Bruneval, Marinopoulos, Dash, Botti,
  Olevano, Vast, Rubio, and Reining}]{QUA:QUA20486}
\bibinfo{author}{F.~Sottile}, \bibinfo{author}{F.~Bruneval},
  \bibinfo{author}{A.~G. Marinopoulos}, \bibinfo{author}{L.~K. Dash},
  \bibinfo{author}{S.~Botti}, \bibinfo{author}{V.~Olevano},
  \bibinfo{author}{N.~Vast}, \bibinfo{author}{A.~Rubio},
  \bibinfo{author}{L.~Reining}, \bibinfo{journal}{Int. J. Quantum Chem.}
  \bibinfo{volume}{102} (\bibinfo{year}{2005}) \bibinfo{pages}{684}.
\bibitem[{Sharma et~al.(2005)Sharma, Dewhurst, and
  Ambrosch-Draxl}]{PhysRevLett.95.136402}
\bibinfo{author}{S.~Sharma}, \bibinfo{author}{J.~K. Dewhurst},
  \bibinfo{author}{C.~Ambrosch-Draxl}, \bibinfo{journal}{Phys. Rev. Lett.}
  \bibinfo{volume}{95} (\bibinfo{year}{2005}) \bibinfo{pages}{136402}.
\bibitem[{Sagmeister and Ambrosch-Draxl(2009)}]{B903676H}
\bibinfo{author}{S.~Sagmeister}, \bibinfo{author}{C.~Ambrosch-Draxl},
  \bibinfo{journal}{Phys. Chem. Chem. Phys.} \bibinfo{volume}{11}
  (\bibinfo{year}{2009}) \bibinfo{pages}{4451}.
\bibitem[{Mortensen et~al.(2005)Mortensen, Hansen, and
  Jacobsen}]{PhysRevB.71.035109}
\bibinfo{author}{J.~J. Mortensen}, \bibinfo{author}{L.~B. Hansen},
  \bibinfo{author}{K.~W. Jacobsen}, \bibinfo{journal}{Phys. Rev. B}
  \bibinfo{volume}{71} (\bibinfo{year}{2005}) \bibinfo{pages}{035109}.
\bibitem[{Enkovaara et~al.(2010)Enkovaara, Rostgaard, Mortensen, Chen, Dułak,
  Ferrighi, Gavnholt, Glinsvad, Haikola, Hansen, Kristoffersen, Kuisma, Larsen,
  Lehtovaara, Ljungberg, Lopez-Acevedo, Moses, Ojanen, Olsen, Petzold, Romero,
  Stausholm-Møller, Strange, Tritsaris, Vanin, Walter, Hammer, Häkkinen,
  Madsen, Nieminen, Nørskov, Puska, Rantala, Schiøtz, Thygesen, and
  Jacobsen}]{0953-8984-22-25-253202}
\bibinfo{author}{J.~Enkovaara}, \bibinfo{author}{C.~Rostgaard},
  \bibinfo{author}{J.~J. Mortensen}, \bibinfo{author}{J.~Chen},
  \bibinfo{author}{M.~Dułak}, \bibinfo{author}{L.~Ferrighi},
  \bibinfo{author}{J.~Gavnholt}, \bibinfo{author}{C.~Glinsvad},
  \bibinfo{author}{V.~Haikola}, \bibinfo{author}{H.~A. Hansen},
  \bibinfo{author}{H.~H. Kristoffersen}, \bibinfo{author}{M.~Kuisma},
  \bibinfo{author}{A.~H. Larsen}, \bibinfo{author}{L.~Lehtovaara},
  \bibinfo{author}{M.~Ljungberg}, \bibinfo{author}{O.~Lopez-Acevedo},
  \bibinfo{author}{P.~G. Moses}, \bibinfo{author}{J.~Ojanen},
  \bibinfo{author}{T.~Olsen}, \bibinfo{author}{V.~Petzold},
  \bibinfo{author}{N.~A. Romero}, \bibinfo{author}{J.~Stausholm-Møller},
  \bibinfo{author}{M.~Strange}, \bibinfo{author}{G.~A. Tritsaris},
  \bibinfo{author}{M.~Vanin}, \bibinfo{author}{M.~Walter},
  \bibinfo{author}{B.~Hammer}, \bibinfo{author}{H.~Häkkinen},
  \bibinfo{author}{G.~K.~H. Madsen}, \bibinfo{author}{R.~M. Nieminen},
  \bibinfo{author}{J.~K. Nørskov}, \bibinfo{author}{M.~Puska},
  \bibinfo{author}{T.~T. Rantala}, \bibinfo{author}{J.~Schiøtz},
  \bibinfo{author}{K.~S. Thygesen}, \bibinfo{author}{K.~W. Jacobsen},
  \bibinfo{journal}{J. Phys.: Cond. Matt.} \bibinfo{volume}{22}
  (\bibinfo{year}{2010}) \bibinfo{pages}{253202}.
\bibitem[{Lathiotakis and Marques(2008)}]{lathiotakis:184103}
\bibinfo{author}{N.~N. Lathiotakis}, \bibinfo{author}{M.~A.~L. Marques},
  \bibinfo{journal}{J. Chem. Phys.} \bibinfo{volume}{128}
  (\bibinfo{year}{2008}) \bibinfo{pages}{184103}.
\bibitem[{Helbig et~al.(2007)Helbig, Lathiotakis, Albrecht, and
  Gross}]{0295-5075-77-6-67003}
\bibinfo{author}{N.~Helbig}, \bibinfo{author}{N.~N. Lathiotakis},
  \bibinfo{author}{M.~Albrecht}, \bibinfo{author}{E.~K.~U. Gross},
  \bibinfo{journal}{Europhys. Lett.} \bibinfo{volume}{77}
  (\bibinfo{year}{2007}) \bibinfo{pages}{67003}.
\bibitem[{Marques and Lathiotakis(2008)}]{PhysRevA.77.032509}
\bibinfo{author}{M.~A.~L. Marques}, \bibinfo{author}{N.~N. Lathiotakis},
  \bibinfo{journal}{Phys. Rev. A} \bibinfo{volume}{77} (\bibinfo{year}{2008})
  \bibinfo{pages}{032509}.
\bibitem[{Marini et~al.(2009)Marini, Hogan, Grüning, and
  Varsano}]{Marini20091392}
\bibinfo{author}{A.~Marini}, \bibinfo{author}{C.~Hogan},
  \bibinfo{author}{M.~Grüning}, \bibinfo{author}{D.~Varsano},
  \bibinfo{journal}{Comp. Phys. Comm.} \bibinfo{volume}{180}
  (\bibinfo{year}{2009}) \bibinfo{pages}{1392}.
\bibitem[{Giannozzi et~al.(2009)Giannozzi, Baroni, Bonini, Calandra, Car,
  Cavazzoni, Ceresoli, Chiarotti, Cococcioni, Dabo, {Dal Corso}, de~Gironcoli,
  Fabris, Fratesi, Gebauer, Gerstmann, Gougoussis, Kokalj, Lazzeri,
  Martin-Samos, Marzari, Mauri, Mazzarello, Paolini, Pasquarello, Paulatto,
  Sbraccia, Scandolo, Sclauzero, Seitsonen, Smogunov, Umari, and
  Wentzcovitch}]{QE-2009}
\bibinfo{author}{P.~Giannozzi}, \bibinfo{author}{S.~Baroni},
  \bibinfo{author}{N.~Bonini}, \bibinfo{author}{M.~Calandra},
  \bibinfo{author}{R.~Car}, \bibinfo{author}{C.~Cavazzoni},
  \bibinfo{author}{D.~Ceresoli}, \bibinfo{author}{G.~L. Chiarotti},
  \bibinfo{author}{M.~Cococcioni}, \bibinfo{author}{I.~Dabo},
  \bibinfo{author}{A.~{Dal Corso}}, \bibinfo{author}{S.~de~Gironcoli},
  \bibinfo{author}{S.~Fabris}, \bibinfo{author}{G.~Fratesi},
  \bibinfo{author}{R.~Gebauer}, \bibinfo{author}{U.~Gerstmann},
  \bibinfo{author}{C.~Gougoussis}, \bibinfo{author}{A.~Kokalj},
  \bibinfo{author}{M.~Lazzeri}, \bibinfo{author}{L.~Martin-Samos},
  \bibinfo{author}{N.~Marzari}, \bibinfo{author}{F.~Mauri},
  \bibinfo{author}{R.~Mazzarello}, \bibinfo{author}{S.~Paolini},
  \bibinfo{author}{A.~Pasquarello}, \bibinfo{author}{L.~Paulatto},
  \bibinfo{author}{C.~Sbraccia}, \bibinfo{author}{S.~Scandolo},
  \bibinfo{author}{G.~Sclauzero}, \bibinfo{author}{A.~P. Seitsonen},
  \bibinfo{author}{A.~Smogunov}, \bibinfo{author}{P.~Umari},
  \bibinfo{author}{R.~M. Wentzcovitch}, \bibinfo{journal}{J. Phys.: Cond.
  Matt.} \bibinfo{volume}{21} (\bibinfo{year}{2009}) \bibinfo{pages}{395502}.
\bibitem[{Dirac(1930)}]{CambridgeJournals:2040328}
\bibinfo{author}{P.~A.~M. Dirac}, \bibinfo{journal}{Math. Proc. Cambridge
  Philos. Soc.} \bibinfo{volume}{26} (\bibinfo{year}{1930})
  \bibinfo{pages}{376}.
\bibitem[{Bloch(1929)}]{springerlink:10.1007/BF01340281}
\bibinfo{author}{F.~Bloch}, \bibinfo{journal}{Z. Phys. A: Hadrons Nucl.}
  \bibinfo{volume}{57} (\bibinfo{year}{1929}) \bibinfo{pages}{545}.
\bibitem[{Helbig et~al.(2011)Helbig, Fuks, Casula, Verstraete, Marques,
  Tokatly, and Rubio}]{PhysRevA.83.032503}
\bibinfo{author}{N.~Helbig}, \bibinfo{author}{J.~I. Fuks},
  \bibinfo{author}{M.~Casula}, \bibinfo{author}{M.~J. Verstraete},
  \bibinfo{author}{M.~A.~L. Marques}, \bibinfo{author}{I.~V. Tokatly},
  \bibinfo{author}{A.~Rubio}, \bibinfo{journal}{Phys. Rev. A}
  \bibinfo{volume}{83} (\bibinfo{year}{2011}) \bibinfo{pages}{032503}.
\bibitem[{Wigner(1938)}]{TF9383400678}
\bibinfo{author}{E.~Wigner}, \bibinfo{journal}{Trans. Faraday Soc.}
  \bibinfo{volume}{34} (\bibinfo{year}{1938}) \bibinfo{pages}{678}.
\bibitem[{Gell-Mann and Brueckner(1957)}]{PhysRev.106.364}
\bibinfo{author}{M.~Gell-Mann}, \bibinfo{author}{K.~A. Brueckner},
  \bibinfo{journal}{Phys. Rev.} \bibinfo{volume}{106} (\bibinfo{year}{1957})
  \bibinfo{pages}{364}.
\bibitem[{Hedin and Lundqvist(1971)}]{0022-3719-4-14-022}
\bibinfo{author}{L.~Hedin}, \bibinfo{author}{B.~I. Lundqvist},
  \bibinfo{journal}{J. Phys. C: Solid State Phys.} \bibinfo{volume}{4}
  (\bibinfo{year}{1971}) \bibinfo{pages}{2064}.
\bibitem[{Gunnarsson and Lundqvist(1976)}]{PhysRevB.13.4274}
\bibinfo{author}{O.~Gunnarsson}, \bibinfo{author}{B.~I. Lundqvist},
  \bibinfo{journal}{Phys. Rev. B} \bibinfo{volume}{13} (\bibinfo{year}{1976})
  \bibinfo{pages}{4274}.
\bibitem[{Vosko et~al.(1980)Vosko, Wilk, and Nusair}]{vwn1980}
\bibinfo{author}{S.~H. Vosko}, \bibinfo{author}{L.~Wilk},
  \bibinfo{author}{M.~Nusair}, \bibinfo{journal}{Can. J. Phys.}
  \bibinfo{volume}{58} (\bibinfo{year}{1980}) \bibinfo{pages}{1200}.
\bibitem[{Perdew and Zunger(1981)}]{PhysRevB.23.5048}
\bibinfo{author}{J.~P. Perdew}, \bibinfo{author}{A.~Zunger},
  \bibinfo{journal}{Phys. Rev. B} \bibinfo{volume}{23} (\bibinfo{year}{1981})
  \bibinfo{pages}{5048}.
\bibitem[{Ortiz and Ballone(1997)}]{PhysRevB.56.9970}
\bibinfo{author}{G.~Ortiz}, \bibinfo{author}{P.~Ballone},
  \bibinfo{journal}{Phys. Rev. B} \bibinfo{volume}{56} (\bibinfo{year}{1997})
  \bibinfo{pages}{9970}.
\bibitem[{Perdew and Wang(1992)}]{PhysRevB.45.13244}
\bibinfo{author}{J.~P. Perdew}, \bibinfo{author}{Y.~Wang},
  \bibinfo{journal}{Phys. Rev. B} \bibinfo{volume}{45} (\bibinfo{year}{1992})
  \bibinfo{pages}{13244}.
\bibitem[{Attaccalite et~al.(2002)Attaccalite, Moroni, Gori-Giorgi, and
  Bachelet}]{PhysRevLett.88.256601}
\bibinfo{author}{C.~Attaccalite}, \bibinfo{author}{S.~Moroni},
  \bibinfo{author}{P.~Gori-Giorgi}, \bibinfo{author}{G.~B. Bachelet},
  \bibinfo{journal}{Phys. Rev. Lett.} \bibinfo{volume}{88}
  (\bibinfo{year}{2002}) \bibinfo{pages}{256601}.
\bibitem[{Pittalis et~al.(2008)Pittalis, R\"as\"anen, and
  Marques}]{PhysRevB.78.195322}
\bibinfo{author}{S.~Pittalis}, \bibinfo{author}{E.~R\"as\"anen},
  \bibinfo{author}{M.~A.~L. Marques}, \bibinfo{journal}{Phys. Rev. B}
  \bibinfo{volume}{78} (\bibinfo{year}{2008}) \bibinfo{pages}{195322}.
\bibitem[{von Barth and Hedin(1972)}]{0022-3719-5-13-012}
\bibinfo{author}{U.~von Barth}, \bibinfo{author}{L.~Hedin},
  \bibinfo{journal}{J. Phys. C: Solid State Phys.} \bibinfo{volume}{5}
  (\bibinfo{year}{1972}) \bibinfo{pages}{1629}.
\bibitem[{Casula et~al.(2006)Casula, Sorella, and
  Senatore}]{PhysRevB.74.245427}
\bibinfo{author}{M.~Casula}, \bibinfo{author}{S.~Sorella},
  \bibinfo{author}{G.~Senatore}, \bibinfo{journal}{Phys. Rev. B}
  \bibinfo{volume}{74} (\bibinfo{year}{2006}) \bibinfo{pages}{245427}.
\bibitem[{Proynov and Salahub(1994)}]{PhysRevB.49.7874}
\bibinfo{author}{E.~I. Proynov}, \bibinfo{author}{D.~R. Salahub},
  \bibinfo{journal}{Phys. Rev. B} \bibinfo{volume}{49} (\bibinfo{year}{1994})
  \bibinfo{pages}{7874}.
\bibitem[{Gomb\'as(1967)}]{gombas}
\bibinfo{author}{P.~Gomb\'as}, \bibinfo{title}{Pseudopotentiale},
  \bibinfo{publisher}{Springer}, \bibinfo{address}{New York},
  \bibinfo{year}{1967}.
\bibitem[{Goedecker et~al.(1996)Goedecker, Teter, and
  Hutter}]{PhysRevB.54.1703}
\bibinfo{author}{S.~Goedecker}, \bibinfo{author}{M.~Teter},
  \bibinfo{author}{J.~Hutter}, \bibinfo{journal}{Phys. Rev. B}
  \bibinfo{volume}{54} (\bibinfo{year}{1996}) \bibinfo{pages}{1703}.
\bibitem[{Lee and Parr(1987)}]{PhysRevA.35.2377}
\bibinfo{author}{C.~Lee}, \bibinfo{author}{R.~G. Parr}, \bibinfo{journal}{Phys.
  Rev. A} \bibinfo{volume}{35} (\bibinfo{year}{1987}) \bibinfo{pages}{2377}.
\bibitem[{Perdew et~al.(1996)Perdew, Burke, and
  Ernzerhof}]{PhysRevLett.77.3865}
\bibinfo{author}{J.~P. Perdew}, \bibinfo{author}{K.~Burke},
  \bibinfo{author}{M.~Ernzerhof}, \bibinfo{journal}{Phys. Rev. Lett.}
  \bibinfo{volume}{77} (\bibinfo{year}{1996}) \bibinfo{pages}{3865}.
\bibitem[{Perdew et~al.(1997)Perdew, Burke, and
  Ernzerhof}]{PhysRevLett.78.1396}
\bibinfo{author}{J.~P. Perdew}, \bibinfo{author}{K.~Burke},
  \bibinfo{author}{M.~Ernzerhof}, \bibinfo{journal}{Phys. Rev. Lett.}
  \bibinfo{volume}{78} (\bibinfo{year}{1997}) \bibinfo{pages}{1396}.
\bibitem[{Zhang and Yang(1998)}]{PhysRevLett.80.890}
\bibinfo{author}{Y.~Zhang}, \bibinfo{author}{W.~Yang}, \bibinfo{journal}{Phys.
  Rev. Lett.} \bibinfo{volume}{80} (\bibinfo{year}{1998}) \bibinfo{pages}{890}.
\bibitem[{Adamo and Barone(2002)}]{adamo:5933}
\bibinfo{author}{C.~Adamo}, \bibinfo{author}{V.~Barone}, \bibinfo{journal}{J.
  Chem. Phys.} \bibinfo{volume}{116} (\bibinfo{year}{2002})
  \bibinfo{pages}{5933}.
\bibitem[{Xu and Goddard(2004)}]{xu:4068}
\bibinfo{author}{X.~Xu}, \bibinfo{author}{W.~A. Goddard}, \bibinfo{journal}{J.
  Chem. Phys.} \bibinfo{volume}{121} (\bibinfo{year}{2004})
  \bibinfo{pages}{4068}.
\bibitem[{Becke(1986)}]{becke:4524}
\bibinfo{author}{A.~D. Becke}, \bibinfo{journal}{J. Chem. Phys.}
  \bibinfo{volume}{84} (\bibinfo{year}{1986}) \bibinfo{pages}{4524}.
\bibitem[{Becke(1997)}]{becke:8554}
\bibinfo{author}{A.~D. Becke}, \bibinfo{journal}{J. Chem. Phys.}
  \bibinfo{volume}{107} (\bibinfo{year}{1997}) \bibinfo{pages}{8554}.
\bibitem[{Becke(1986)}]{becke:7184}
\bibinfo{author}{A.~D. Becke}, \bibinfo{journal}{J. Chem. Phys.}
  \bibinfo{volume}{85} (\bibinfo{year}{1986}) \bibinfo{pages}{7184}.
\bibitem[{Becke(1988)}]{PhysRevA.38.3098}
\bibinfo{author}{A.~D. Becke}, \bibinfo{journal}{Phys. Rev. A}
  \bibinfo{volume}{38} (\bibinfo{year}{1988}) \bibinfo{pages}{3098}.
\bibitem[{Gill(1996)}]{doi:10.1080/002689796173813}
\bibinfo{author}{P.~M.~W. Gill}, \bibinfo{journal}{Mol. Phys.}
  \bibinfo{volume}{89} (\bibinfo{year}{1996}) \bibinfo{pages}{433}.
\bibitem[{Perdew and Yue(1986)}]{PhysRevB.33.8800}
\bibinfo{author}{J.~P. Perdew}, \bibinfo{author}{W.~Yue},
  \bibinfo{journal}{Phys. Rev. B} \bibinfo{volume}{33} (\bibinfo{year}{1986})
  \bibinfo{pages}{8800}.
\bibitem[{Perdew et~al.(1992)Perdew, Chevary, Vosko, Jackson, Pederson, Singh,
  and Fiolhais}]{PhysRevB.46.6671}
\bibinfo{author}{J.~P. Perdew}, \bibinfo{author}{J.~A. Chevary},
  \bibinfo{author}{S.~H. Vosko}, \bibinfo{author}{K.~A. Jackson},
  \bibinfo{author}{M.~R. Pederson}, \bibinfo{author}{D.~J. Singh},
  \bibinfo{author}{C.~Fiolhais}, \bibinfo{journal}{Phys. Rev. B}
  \bibinfo{volume}{46} (\bibinfo{year}{1992}) \bibinfo{pages}{6671}.
\bibitem[{Handy and Cohen(2001)}]{doi:10.1080/00268970010018431}
\bibinfo{author}{N.~C. Handy}, \bibinfo{author}{A.~J. Cohen},
  \bibinfo{journal}{Mol. Phys.} \bibinfo{volume}{99} (\bibinfo{year}{2001})
  \bibinfo{pages}{403}.
\bibitem[{DePristo and Kress(1987)}]{depristo:1425}
\bibinfo{author}{A.~E. DePristo}, \bibinfo{author}{J.~D. Kress},
  \bibinfo{journal}{J. Chem. Phys.} \bibinfo{volume}{86} (\bibinfo{year}{1987})
  \bibinfo{pages}{1425}.
\bibitem[{Lacks and Gordon(1993)}]{PhysRevA.47.4681}
\bibinfo{author}{D.~J. Lacks}, \bibinfo{author}{R.~G. Gordon},
  \bibinfo{journal}{Phys. Rev. A} \bibinfo{volume}{47} (\bibinfo{year}{1993})
  \bibinfo{pages}{4681}.
\bibitem[{Filatov and Thiel(1997)}]{doi:10.1080/002689797170950}
\bibinfo{author}{M.~Filatov}, \bibinfo{author}{W.~Thiel},
  \bibinfo{journal}{Mol. Phys.} \bibinfo{volume}{91} (\bibinfo{year}{1997})
  \bibinfo{pages}{847}.
\bibitem[{Perdew et~al.(2008)Perdew, Ruzsinszky, Csonka, Vydrov, Scuseria,
  Constantin, Zhou, and Burke}]{PhysRevLett.100.136406}
\bibinfo{author}{J.~P. Perdew}, \bibinfo{author}{A.~Ruzsinszky},
  \bibinfo{author}{G.~I. Csonka}, \bibinfo{author}{O.~A. Vydrov},
  \bibinfo{author}{G.~E. Scuseria}, \bibinfo{author}{L.~A. Constantin},
  \bibinfo{author}{X.~Zhou}, \bibinfo{author}{K.~Burke},
  \bibinfo{journal}{Phys. Rev. Lett.} \bibinfo{volume}{100}
  (\bibinfo{year}{2008}) \bibinfo{pages}{136406}.
\bibitem[{Hammer et~al.(1999)Hammer, Hansen, and N\o{}rskov}]{PhysRevB.59.7413}
\bibinfo{author}{B.~Hammer}, \bibinfo{author}{L.~B. Hansen},
  \bibinfo{author}{J.~K. N\o{}rskov}, \bibinfo{journal}{Phys. Rev. B}
  \bibinfo{volume}{59} (\bibinfo{year}{1999}) \bibinfo{pages}{7413}.
\bibitem[{Wu and Cohen(2006)}]{PhysRevB.73.235116}
\bibinfo{author}{Z.~Wu}, \bibinfo{author}{R.~E. Cohen}, \bibinfo{journal}{Phys.
  Rev. B} \bibinfo{volume}{73} (\bibinfo{year}{2006}) \bibinfo{pages}{235116}.
\bibitem[{Armiento and Mattsson(2005)}]{PhysRevB.72.085108}
\bibinfo{author}{R.~Armiento}, \bibinfo{author}{A.~E. Mattsson},
  \bibinfo{journal}{Phys. Rev. B} \bibinfo{volume}{72} (\bibinfo{year}{2005})
  \bibinfo{pages}{085108}.
\bibitem[{Mattsson et~al.(2008)Mattsson, Armiento, Paier, Kresse, Wills, and
  Mattsson}]{mattsson:084714}
\bibinfo{author}{A.~E. Mattsson}, \bibinfo{author}{R.~Armiento},
  \bibinfo{author}{J.~Paier}, \bibinfo{author}{G.~Kresse},
  \bibinfo{author}{J.~M. Wills}, \bibinfo{author}{T.~R. Mattsson},
  \bibinfo{journal}{J. Chem. Phys.} \bibinfo{volume}{128}
  (\bibinfo{year}{2008}) \bibinfo{pages}{084714}.
\bibitem[{Madsen(2007)}]{PhysRevB.75.195108}
\bibinfo{author}{G.~K.~H. Madsen}, \bibinfo{journal}{Phys. Rev. B}
  \bibinfo{volume}{75} (\bibinfo{year}{2007}) \bibinfo{pages}{195108}.
\bibitem[{Adamo and Barone(1998)}]{adamo:664}
\bibinfo{author}{C.~Adamo}, \bibinfo{author}{V.~Barone}, \bibinfo{journal}{J.
  Chem. Phys.} \bibinfo{volume}{108} (\bibinfo{year}{1998})
  \bibinfo{pages}{664}.
\bibitem[{Pittalis et~al.(2009)Pittalis, R\"as\"anen, Vilhena, and
  Marques}]{PhysRevA.79.012503}
\bibinfo{author}{S.~Pittalis}, \bibinfo{author}{E.~R\"as\"anen},
  \bibinfo{author}{J.~G. Vilhena}, \bibinfo{author}{M.~A.~L. Marques},
  \bibinfo{journal}{Phys. Rev. A} \bibinfo{volume}{79} (\bibinfo{year}{2009})
  \bibinfo{pages}{012503}.
\bibitem[{Mortensen et~al.(2005)Mortensen, Kaasbjerg, Frederiksen, N\o{}rskov,
  Sethna, and Jacobsen}]{PhysRevLett.95.216401}
\bibinfo{author}{J.~J. Mortensen}, \bibinfo{author}{K.~Kaasbjerg},
  \bibinfo{author}{S.~L. Frederiksen}, \bibinfo{author}{J.~K. N\o{}rskov},
  \bibinfo{author}{J.~P. Sethna}, \bibinfo{author}{K.~W. Jacobsen},
  \bibinfo{journal}{Phys. Rev. Lett.} \bibinfo{volume}{95}
  (\bibinfo{year}{2005}) \bibinfo{pages}{216401}.
\bibitem[{Pedroza et~al.(2009)Pedroza, da~Silva, and
  Capelle}]{PhysRevB.79.201106}
\bibinfo{author}{L.~S. Pedroza}, \bibinfo{author}{A.~J.~R. da~Silva},
  \bibinfo{author}{K.~Capelle}, \bibinfo{journal}{Phys. Rev. B}
  \bibinfo{volume}{79} (\bibinfo{year}{2009}) \bibinfo{pages}{201106}.
\bibitem[{Klime\v{s} et~al.(2010)Klime\v{s}, Bowler, and
  Michaelides}]{0953-8984-22-2-022201}
\bibinfo{author}{J.~Klime\v{s}}, \bibinfo{author}{D.~R. Bowler},
  \bibinfo{author}{A.~Michaelides}, \bibinfo{journal}{J. Phys.: Cond. Matt.}
  \bibinfo{volume}{22} (\bibinfo{year}{2010}) \bibinfo{pages}{022201}.
\bibitem[{Ruzsinszky et~al.(2009)Ruzsinszky, Csonka, and
  Scuseria}]{doi:10.1021/ct8005369}
\bibinfo{author}{A.~Ruzsinszky}, \bibinfo{author}{G.~I. Csonka},
  \bibinfo{author}{G.~E. Scuseria}, \bibinfo{journal}{J. Chem. Theo. Comp.}
  \bibinfo{volume}{5} (\bibinfo{year}{2009}) \bibinfo{pages}{763}.
\bibitem[{Murray et~al.(2009)Murray, Lee, and Langreth}]{doi:10.1021/ct900365q}
\bibinfo{author}{{\'E}.~D. Murray}, \bibinfo{author}{K.~Lee},
  \bibinfo{author}{D.~C. Langreth}, \bibinfo{journal}{J. Chem. Theory Comput.}
  \bibinfo{volume}{5} (\bibinfo{year}{2009}) \bibinfo{pages}{2754}.
\bibitem[{Keal and Tozer(2003)}]{keal:3015}
\bibinfo{author}{T.~W. Keal}, \bibinfo{author}{D.~J. Tozer},
  \bibinfo{journal}{J. Chem. Phys.} \bibinfo{volume}{119}
  (\bibinfo{year}{2003}) \bibinfo{pages}{3015}.
\bibitem[{Herman et~al.(1969{\natexlab{a}})Herman, Van~Dyke, and
  Ortenburger}]{PhysRevLett.22.807}
\bibinfo{author}{F.~Herman}, \bibinfo{author}{J.~P. Van~Dyke},
  \bibinfo{author}{I.~B. Ortenburger}, \bibinfo{journal}{Phys. Rev. Lett.}
  \bibinfo{volume}{22} (\bibinfo{year}{1969}{\natexlab{a}})
  \bibinfo{pages}{807}.
\bibitem[{Herman et~al.(1969{\natexlab{b}})Herman, Ortenburger, and
  Van~Dyke}]{QUA:QUA560040746}
\bibinfo{author}{F.~Herman}, \bibinfo{author}{I.~B. Ortenburger},
  \bibinfo{author}{J.~P. Van~Dyke}, \bibinfo{journal}{Int. J. Quant. Chem.}
  \bibinfo{volume}{4} (\bibinfo{year}{1969}{\natexlab{b}})
  \bibinfo{pages}{827}.
\bibitem[{Schipper et~al.(2000)Schipper, Gritsenko, van Gisbergen, and
  Baerends}]{schipper:1344}
\bibinfo{author}{P.~R.~T. Schipper}, \bibinfo{author}{O.~V. Gritsenko},
  \bibinfo{author}{S.~J.~A. van Gisbergen}, \bibinfo{author}{E.~J. Baerends},
  \bibinfo{journal}{J. Chem. Phys.} \bibinfo{volume}{112}
  (\bibinfo{year}{2000}) \bibinfo{pages}{1344}.
\bibitem[{Fuentealba and Reyes(1995)}]{Fuentealba199531}
\bibinfo{author}{P.~Fuentealba}, \bibinfo{author}{O.~Reyes},
  \bibinfo{journal}{Chem. Phys. Lett.} \bibinfo{volume}{232}
  (\bibinfo{year}{1995}) \bibinfo{pages}{31}.
\bibitem[{Ou-Yang and Levy(1991)}]{QUA:QUA560400309}
\bibinfo{author}{H.~Ou-Yang}, \bibinfo{author}{M.~Levy}, \bibinfo{journal}{Int.
  J. Quantum Chem.} \bibinfo{volume}{40} (\bibinfo{year}{1991})
  \bibinfo{pages}{379}.
\bibitem[{Tognetti and Adamo(2009)}]{doi:10.1021/jp903672e}
\bibinfo{author}{V.~Tognetti}, \bibinfo{author}{C.~Adamo}, \bibinfo{journal}{J.
  Phys. Chem. A} \bibinfo{volume}{113} (\bibinfo{year}{2009})
  \bibinfo{pages}{14415}. \bibinfo{note}{PMID: 19518066}.
\bibitem[{Constantin et~al.(2011)Constantin, Fabiano, Laricchia, and
  Della~Sala}]{PhysRevLett.106.186406}
\bibinfo{author}{L.~A. Constantin}, \bibinfo{author}{E.~Fabiano},
  \bibinfo{author}{S.~Laricchia}, \bibinfo{author}{F.~Della~Sala},
  \bibinfo{journal}{Phys. Rev. Lett.} \bibinfo{volume}{106}
  (\bibinfo{year}{2011}) \bibinfo{pages}{186406}.
\bibitem[{Haas et~al.(2011)Haas, Tran, Blaha, and Schwarz}]{PhysRevB.83.205117}
\bibinfo{author}{P.~Haas}, \bibinfo{author}{F.~Tran},
  \bibinfo{author}{P.~Blaha}, \bibinfo{author}{K.~Schwarz},
  \bibinfo{journal}{Phys. Rev. B} \bibinfo{volume}{83} (\bibinfo{year}{2011})
  \bibinfo{pages}{205117}.
\bibitem[{Constantin et~al.(2009)Constantin, Ruzsinszky, and
  Perdew}]{PhysRevB.80.035125}
\bibinfo{author}{L.~A. Constantin}, \bibinfo{author}{A.~Ruzsinszky},
  \bibinfo{author}{J.~P. Perdew}, \bibinfo{journal}{Phys. Rev. B}
  \bibinfo{volume}{80} (\bibinfo{year}{2009}) \bibinfo{pages}{035125}.
\bibitem[{Vitos et~al.(2000)Vitos, Johansson, Koll\'ar, and
  Skriver}]{PhysRevB.62.10046}
\bibinfo{author}{L.~Vitos}, \bibinfo{author}{B.~Johansson},
  \bibinfo{author}{J.~Koll\'ar}, \bibinfo{author}{H.~L. Skriver},
  \bibinfo{journal}{Phys. Rev. B} \bibinfo{volume}{62} (\bibinfo{year}{2000})
  \bibinfo{pages}{10046}.
\bibitem[{Peverati et~al.(2011)Peverati, Zhao, and
  Truhlar}]{doi:10.1021/jz200616w}
\bibinfo{author}{R.~Peverati}, \bibinfo{author}{Y.~Zhao},
  \bibinfo{author}{D.~G. Truhlar}, \bibinfo{journal}{J. Phys. Chem. Lett.}
  \bibinfo{volume}{2} (\bibinfo{year}{2011}) \bibinfo{pages}{1991}.
\bibitem[{Cooper(2010)}]{PhysRevB.81.161104}
\bibinfo{author}{V.~R. Cooper}, \bibinfo{journal}{Phys. Rev. B}
  \bibinfo{volume}{81} (\bibinfo{year}{2010}) \bibinfo{pages}{161104}.
\bibitem[{Lee et~al.(1988)Lee, Yang, and Parr}]{PhysRevB.37.785}
\bibinfo{author}{C.~Lee}, \bibinfo{author}{W.~Yang}, \bibinfo{author}{R.~G.
  Parr}, \bibinfo{journal}{Phys. Rev. B} \bibinfo{volume}{37}
  (\bibinfo{year}{1988}) \bibinfo{pages}{785}.
\bibitem[{Miehlich et~al.(1989)Miehlich, Savin, Stoll, and
  Preuss}]{Miehlich1989200}
\bibinfo{author}{B.~Miehlich}, \bibinfo{author}{A.~Savin},
  \bibinfo{author}{H.~Stoll}, \bibinfo{author}{H.~Preuss},
  \bibinfo{journal}{Chem. Phys. Lett.} \bibinfo{volume}{157}
  (\bibinfo{year}{1989}) \bibinfo{pages}{200}.
\bibitem[{Perdew(1986)}]{PhysRevB.33.8822}
\bibinfo{author}{J.~P. Perdew}, \bibinfo{journal}{Phys. Rev. B}
  \bibinfo{volume}{33} (\bibinfo{year}{1986}) \bibinfo{pages}{8822}.
\bibitem[{Perdew et~al.(1993)Perdew, Chevary, Vosko, Jackson, Pederson, Singh,
  and Fiolhais}]{PhysRevB.48.4978.2}
\bibinfo{author}{J.~P. Perdew}, \bibinfo{author}{J.~A. Chevary},
  \bibinfo{author}{S.~H. Vosko}, \bibinfo{author}{K.~A. Jackson},
  \bibinfo{author}{M.~R. Pederson}, \bibinfo{author}{D.~J. Singh},
  \bibinfo{author}{C.~Fiolhais}, \bibinfo{journal}{Phys. Rev. B}
  \bibinfo{volume}{48} (\bibinfo{year}{1993}) \bibinfo{pages}{4978}.
\bibitem[{Langreth and Mehl(1981)}]{PhysRevLett.47.446}
\bibinfo{author}{D.~C. Langreth}, \bibinfo{author}{M.~J. Mehl},
  \bibinfo{journal}{Phys. Rev. Lett.} \bibinfo{volume}{47}
  (\bibinfo{year}{1981}) \bibinfo{pages}{446}.
\bibitem[{Wilson and Levy(1990)}]{PhysRevB.41.12930}
\bibinfo{author}{L.~C. Wilson}, \bibinfo{author}{M.~Levy},
  \bibinfo{journal}{Phys. Rev. B} \bibinfo{volume}{41} (\bibinfo{year}{1990})
  \bibinfo{pages}{12930}.
\bibitem[{Wilson and Ivanov(1998)}]{QUA:QUA9}
\bibinfo{author}{L.~C. Wilson}, \bibinfo{author}{S.~Ivanov},
  \bibinfo{journal}{Int. J. Quant. Chem.} \bibinfo{volume}{69}
  (\bibinfo{year}{1998}) \bibinfo{pages}{523}.
\bibitem[{Peverati and Truhlar(2011)}]{peverati:191102}
\bibinfo{author}{R.~Peverati}, \bibinfo{author}{D.~G. Truhlar},
  \bibinfo{journal}{J. Chem. Phys.} \bibinfo{volume}{135}
  (\bibinfo{year}{2011}) \bibinfo{pages}{191102}.
\bibitem[{van Leeuwen and Baerends(1994)}]{PhysRevA.49.2421}
\bibinfo{author}{R.~van Leeuwen}, \bibinfo{author}{E.~J. Baerends},
  \bibinfo{journal}{Phys. Rev. A} \bibinfo{volume}{49} (\bibinfo{year}{1994})
  \bibinfo{pages}{2421}.
\bibitem[{Hamprecht et~al.(1998)Hamprecht, Cohen, Tozer, and
  Handy}]{hamprecht:6264}
\bibinfo{author}{F.~A. Hamprecht}, \bibinfo{author}{A.~J. Cohen},
  \bibinfo{author}{D.~J. Tozer}, \bibinfo{author}{N.~C. Handy},
  \bibinfo{journal}{J. Chem. Phys.} \bibinfo{volume}{109}
  (\bibinfo{year}{1998}) \bibinfo{pages}{6264}.
\bibitem[{Boese et~al.(2000)Boese, Doltsinis, Handy, and Sprik}]{boese:1670}
\bibinfo{author}{A.~D. Boese}, \bibinfo{author}{N.~L. Doltsinis},
  \bibinfo{author}{N.~C. Handy}, \bibinfo{author}{M.~Sprik},
  \bibinfo{journal}{J. Chem. Phys.} \bibinfo{volume}{112}
  (\bibinfo{year}{2000}) \bibinfo{pages}{1670}.
\bibitem[{Boese and Handy(2001)}]{boese:5497}
\bibinfo{author}{A.~D. Boese}, \bibinfo{author}{N.~C. Handy},
  \bibinfo{journal}{J. Chem. Phys.} \bibinfo{volume}{114}
  (\bibinfo{year}{2001}) \bibinfo{pages}{5497}.
\bibitem[{Adamson et~al.(1998)Adamson, Gill, and Pople}]{Adamson19986}
\bibinfo{author}{R.~D. Adamson}, \bibinfo{author}{P.~M. Gill},
  \bibinfo{author}{J.~A. Pople}, \bibinfo{journal}{Chem. Phys. Lett.}
  \bibinfo{volume}{284} (\bibinfo{year}{1998}) \bibinfo{pages}{6}.
\bibitem[{Xu and Goddard(2004)}]{Xu02032004}
\bibinfo{author}{X.~Xu}, \bibinfo{author}{W.~A. Goddard},
  \bibinfo{journal}{Proc. Nat. Ac. Sci. USA} \bibinfo{volume}{101}
  (\bibinfo{year}{2004}) \bibinfo{pages}{2673}.
\bibitem[{Dahlke and Truhlar(2005)}]{doi:10.1021/jp052436c}
\bibinfo{author}{E.~E. Dahlke}, \bibinfo{author}{D.~G. Truhlar},
  \bibinfo{journal}{J. Phys. Chem. B} \bibinfo{volume}{109}
  (\bibinfo{year}{2005}) \bibinfo{pages}{15677}.
\bibitem[{Tozer et~al.(1997)Tozer, Handy, and Green}]{Tozer1997183}
\bibinfo{author}{D.~J. Tozer}, \bibinfo{author}{N.~C. Handy},
  \bibinfo{author}{W.~H. Green}, \bibinfo{journal}{Chem. Phys. Lett.}
  \bibinfo{volume}{273} (\bibinfo{year}{1997}) \bibinfo{pages}{183}.
\bibitem[{Tozer and Handy(1998{\natexlab{a}})}]{tozer:2545}
\bibinfo{author}{D.~J. Tozer}, \bibinfo{author}{N.~C. Handy},
  \bibinfo{journal}{J. Chem. Phys.} \bibinfo{volume}{108}
  (\bibinfo{year}{1998}{\natexlab{a}}) \bibinfo{pages}{2545}.
\bibitem[{Tozer and Handy(1998{\natexlab{b}})}]{doi:10.1021/jp980259s}
\bibinfo{author}{D.~J. Tozer}, \bibinfo{author}{N.~C. Handy},
  \bibinfo{journal}{J. Phys. Chem. A} \bibinfo{volume}{102}
  (\bibinfo{year}{1998}{\natexlab{b}}) \bibinfo{pages}{3162}.
\bibitem[{Handy and Tozer(1998)}]{doi:10.1080/002689798167863}
\bibinfo{author}{N.~C. Handy}, \bibinfo{author}{D.~J. Tozer},
  \bibinfo{journal}{Mol. Phys.} \bibinfo{volume}{94} (\bibinfo{year}{1998})
  \bibinfo{pages}{707}.
\bibitem[{Weizsäcker(1935)}]{springerlink:10.1007/BF01337700}
\bibinfo{author}{C.~F.~v. Weizsäcker}, \bibinfo{journal}{Z. Phys. A: Hadrons
  Nucl.} \bibinfo{volume}{96} (\bibinfo{year}{1935}) \bibinfo{pages}{431}.
\bibitem[{Kompaneets and Pavlovskii(1956)}]{Kompaneets1956}
\bibinfo{author}{A.~S. Kompaneets}, \bibinfo{author}{E.~S. Pavlovskii},
  \bibinfo{journal}{Zh. Eksp. Teor. Fiz.} \bibinfo{volume}{31}
  (\bibinfo{year}{1956}) \bibinfo{pages}{427}. \bibinfo{note}{[Sov. Phys. JETP
  {\textbf{4}}, 328 (1957)]}.
\bibitem[{Kirznits(1957)}]{Kirznits1957}
\bibinfo{author}{D.~A. Kirznits}, \bibinfo{journal}{Zh. Eksp. Teor. Fiz.}
  \bibinfo{volume}{32} (\bibinfo{year}{1957}) \bibinfo{pages}{115}.
  \bibinfo{note}{[Sov. Phys. JETP {\textbf{5}}, 64 (1957)]}.
\bibitem[{Golden(1957)}]{PhysRev.105.604}
\bibinfo{author}{S.~Golden}, \bibinfo{journal}{Phys. Rev.}
  \bibinfo{volume}{105} (\bibinfo{year}{1957}) \bibinfo{pages}{604}.
\bibitem[{Yonei and Tomishima(1965)}]{JPSJ.20.1051}
\bibinfo{author}{K.~Yonei}, \bibinfo{author}{Y.~Tomishima},
  \bibinfo{journal}{J. Phys. Soc. Jpn.} \bibinfo{volume}{20}
  (\bibinfo{year}{1965}) \bibinfo{pages}{1051}.
\bibitem[{Baltin(1972)}]{Baltin1972}
\bibinfo{author}{R.~Baltin}, \bibinfo{journal}{Z. Naturforsch.}
  \bibinfo{volume}{27} (\bibinfo{year}{1972}) \bibinfo{pages}{1176}.
\bibitem[{Lieb(1981)}]{RevModPhys.53.603}
\bibinfo{author}{E.~H. Lieb}, \bibinfo{journal}{Rev. Mod. Phys.}
  \bibinfo{volume}{53} (\bibinfo{year}{1981}) \bibinfo{pages}{603}.
\bibitem[{Acharya et~al.(1980)Acharya, Bartolotti, Sears, and
  Parr}]{Acharya01121980}
\bibinfo{author}{P.~K. Acharya}, \bibinfo{author}{L.~J. Bartolotti},
  \bibinfo{author}{S.~B. Sears}, \bibinfo{author}{R.~G. Parr},
  \bibinfo{journal}{Proc. Nat. Ac. Sci.} \bibinfo{volume}{77}
  (\bibinfo{year}{1980}) \bibinfo{pages}{6978}.
\bibitem[{G\'{a}zquez and Robles(1982)}]{gazquez:1467}
\bibinfo{author}{J.~L. G\'{a}zquez}, \bibinfo{author}{J.~Robles},
  \bibinfo{journal}{J. Chem. Phys.} \bibinfo{volume}{76} (\bibinfo{year}{1982})
  \bibinfo{pages}{1467}.
\bibitem[{Lude\~na(1986)}]{Ludena1986}
\bibinfo{author}{E.~V. Lude\~na}, in: \bibinfo{editor}{F.~B. Malik} (Ed.),
  \bibinfo{booktitle}{Cond. Matt. Theor.}, volume~\bibinfo{volume}{1},
  \bibinfo{publisher}{Plenum}, \bibinfo{address}{New York},
  \bibinfo{year}{1986}, p. \bibinfo{pages}{183}.
\bibitem[{Ghosh and Parr(1985)}]{ghosh:3307}
\bibinfo{author}{S.~K. Ghosh}, \bibinfo{author}{R.~G. Parr},
  \bibinfo{journal}{J. Chem. Phys.} \bibinfo{volume}{82} (\bibinfo{year}{1985})
  \bibinfo{pages}{3307}.
\bibitem[{Lee et~al.(1991)Lee, Lee, and Parr}]{PhysRevA.44.768}
\bibinfo{author}{H.~Lee}, \bibinfo{author}{C.~Lee}, \bibinfo{author}{R.~G.
  Parr}, \bibinfo{journal}{Phys. Rev. A} \bibinfo{volume}{44}
  (\bibinfo{year}{1991}) \bibinfo{pages}{768}.
\bibitem[{Lacks and Gordon(1994)}]{lacks:4446}
\bibinfo{author}{D.~J. Lacks}, \bibinfo{author}{R.~G. Gordon},
  \bibinfo{journal}{J. Chem. Phys.} \bibinfo{volume}{100}
  (\bibinfo{year}{1994}) \bibinfo{pages}{4446}.
\bibitem[{Pearson and Gordon(1985)}]{pearson:881}
\bibinfo{author}{E.~W. Pearson}, \bibinfo{author}{R.~G. Gordon},
  \bibinfo{journal}{J. Chem. Phys.} \bibinfo{volume}{82} (\bibinfo{year}{1985})
  \bibinfo{pages}{881}.
\bibitem[{DePristo and Kress(1987)}]{PhysRevA.35.438}
\bibinfo{author}{A.~E. DePristo}, \bibinfo{author}{J.~D. Kress},
  \bibinfo{journal}{Phys. Rev. A} \bibinfo{volume}{35} (\bibinfo{year}{1987})
  \bibinfo{pages}{438}.
\bibitem[{P. and Perdew(1992)}]{JohnP199279}
\bibinfo{author}{J.~P.}, \bibinfo{author}{Perdew}, \bibinfo{journal}{Phys.
  Lett. A} \bibinfo{volume}{165} (\bibinfo{year}{1992}) \bibinfo{pages}{79}.
\bibitem[{Vitos et~al.(1998)Vitos, Skriver, and Koll\'ar}]{PhysRevB.57.12611}
\bibinfo{author}{L.~Vitos}, \bibinfo{author}{H.~L. Skriver},
  \bibinfo{author}{J.~Koll\'ar}, \bibinfo{journal}{Phys. Rev. B}
  \bibinfo{volume}{57} (\bibinfo{year}{1998}) \bibinfo{pages}{12611}.
\bibitem[{Vitos et~al.(2000)Vitos, Johansson, Koll\'ar, and
  Skriver}]{PhysRevA.61.052511}
\bibinfo{author}{L.~Vitos}, \bibinfo{author}{B.~Johansson},
  \bibinfo{author}{J.~Koll\'ar}, \bibinfo{author}{H.~L. Skriver},
  \bibinfo{journal}{Phys. Rev. A} \bibinfo{volume}{61} (\bibinfo{year}{2000})
  \bibinfo{pages}{052511}.
\bibitem[{M. and Ernzerhof(2000)}]{M200059}
\bibinfo{author}{M.}, \bibinfo{author}{Ernzerhof}, \bibinfo{journal}{J. Mol.
  Struct.: THEOCHEM} \bibinfo{volume}{501--502} (\bibinfo{year}{2000})
  \bibinfo{pages}{59}.
\bibitem[{Lembarki and Chermette(1994)}]{PhysRevA.50.5328}
\bibinfo{author}{A.~Lembarki}, \bibinfo{author}{H.~Chermette},
  \bibinfo{journal}{Phys. Rev. A} \bibinfo{volume}{50} (\bibinfo{year}{1994})
  \bibinfo{pages}{5328}.
\bibitem[{Thakkar(1992)}]{PhysRevA.46.6920}
\bibinfo{author}{A.~J. Thakkar}, \bibinfo{journal}{Phys. Rev. A}
  \bibinfo{volume}{46} (\bibinfo{year}{1992}) \bibinfo{pages}{6920}.
\bibitem[{Tran and Wesołowski(2002)}]{QUA:QUA10306}
\bibinfo{author}{F.~Tran}, \bibinfo{author}{T.~A. Wesołowski},
  \bibinfo{journal}{Int. J. Quantum Chem.} \bibinfo{volume}{89}
  (\bibinfo{year}{2002}) \bibinfo{pages}{441}.
\bibitem[{Ernzerhof and Scuseria(1999)}]{ernzerhof:911}
\bibinfo{author}{M.~Ernzerhof}, \bibinfo{author}{G.~E. Scuseria},
  \bibinfo{journal}{J. Chem. Phys.} \bibinfo{volume}{111}
  (\bibinfo{year}{1999}) \bibinfo{pages}{911}.
\bibitem[{Tao et~al.(2003)Tao, Perdew, Staroverov, and
  Scuseria}]{PhysRevLett.91.146401}
\bibinfo{author}{J.~Tao}, \bibinfo{author}{J.~P. Perdew},
  \bibinfo{author}{V.~N. Staroverov}, \bibinfo{author}{G.~E. Scuseria},
  \bibinfo{journal}{Phys. Rev. Lett.} \bibinfo{volume}{91}
  (\bibinfo{year}{2003}) \bibinfo{pages}{146401}.
\bibitem[{Perdew et~al.(2004)Perdew, Tao, Staroverov, and
  Scuseria}]{perdew:6898}
\bibinfo{author}{J.~P. Perdew}, \bibinfo{author}{J.~Tao},
  \bibinfo{author}{V.~N. Staroverov}, \bibinfo{author}{G.~E. Scuseria},
  \bibinfo{journal}{J. Chem. Phys.} \bibinfo{volume}{120}
  (\bibinfo{year}{2004}) \bibinfo{pages}{6898}.
\bibitem[{Boese and Handy(2002)}]{boese:9559}
\bibinfo{author}{A.~D. Boese}, \bibinfo{author}{N.~C. Handy},
  \bibinfo{journal}{J. Chem. Phys.} \bibinfo{volume}{116}
  (\bibinfo{year}{2002}) \bibinfo{pages}{9559}.
\bibitem[{Voorhis and Scuseria(1998)}]{voorhis:400}
\bibinfo{author}{T.~V. Voorhis}, \bibinfo{author}{G.~E. Scuseria},
  \bibinfo{journal}{J. Chem. Phys.} \bibinfo{volume}{109}
  (\bibinfo{year}{1998}) \bibinfo{pages}{400}.
\bibitem[{Zhao and Truhlar(2006)}]{zhao:194101}
\bibinfo{author}{Y.~Zhao}, \bibinfo{author}{D.~G. Truhlar},
  \bibinfo{journal}{J. Chem. Phys.} \bibinfo{volume}{125}
  (\bibinfo{year}{2006}) \bibinfo{pages}{194101}.
\bibitem[{Zhao and Truhlar(2008)}]{springerlink:10.1007/s00214-007-0401-8}
\bibinfo{author}{Y.~Zhao}, \bibinfo{author}{D.~Truhlar},
  \bibinfo{journal}{Theor. Chem. Acc.} \bibinfo{volume}{119}
  (\bibinfo{year}{2008}) \bibinfo{pages}{525}.
\bibitem[{Becke and Roussel(1989)}]{PhysRevA.39.3761}
\bibinfo{author}{A.~D. Becke}, \bibinfo{author}{M.~R. Roussel},
  \bibinfo{journal}{Phys. Rev. A} \bibinfo{volume}{39} (\bibinfo{year}{1989})
  \bibinfo{pages}{3761}.
\bibitem[{Becke and Johnson(2006)}]{becke:221101}
\bibinfo{author}{A.~D. Becke}, \bibinfo{author}{E.~R. Johnson},
  \bibinfo{journal}{J. Chem. Phys.} \bibinfo{volume}{124}
  (\bibinfo{year}{2006}) \bibinfo{pages}{221101}.
\bibitem[{Tran and Blaha(2009)}]{PhysRevLett.102.226401}
\bibinfo{author}{F.~Tran}, \bibinfo{author}{P.~Blaha}, \bibinfo{journal}{Phys.
  Rev. Lett.} \bibinfo{volume}{102} (\bibinfo{year}{2009})
  \bibinfo{pages}{226401}.
\bibitem[{R\"{a}s\"{a}nen et~al.(2010)R\"{a}s\"{a}nen, Pittalis, and
  Proetto}]{rasanen:044112}
\bibinfo{author}{E.~R\"{a}s\"{a}nen}, \bibinfo{author}{S.~Pittalis},
  \bibinfo{author}{C.~R. Proetto}, \bibinfo{journal}{J. Chem. Phys.}
  \bibinfo{volume}{132} (\bibinfo{year}{2010}) \bibinfo{pages}{044112}.
\bibitem[{Pittalis et~al.(2007)Pittalis, R\"as\"anen, Helbig, and
  Gross}]{PhysRevB.76.235314}
\bibinfo{author}{S.~Pittalis}, \bibinfo{author}{E.~R\"as\"anen},
  \bibinfo{author}{N.~Helbig}, \bibinfo{author}{E.~K.~U. Gross},
  \bibinfo{journal}{Phys. Rev. B} \bibinfo{volume}{76} (\bibinfo{year}{2007})
  \bibinfo{pages}{235314}.
\bibitem[{Becke(1993)}]{becke:5648}
\bibinfo{author}{A.~D. Becke}, \bibinfo{journal}{J. Chem. Phys.}
  \bibinfo{volume}{98} (\bibinfo{year}{1993}) \bibinfo{pages}{5648}.
\bibitem[{Stephens et~al.(1994)Stephens, Devlin, Chabalowski, and
  Frisch}]{doi:10.1021/j100096a001}
\bibinfo{author}{P.~J. Stephens}, \bibinfo{author}{F.~J. Devlin},
  \bibinfo{author}{C.~F. Chabalowski}, \bibinfo{author}{M.~J. Frisch},
  \bibinfo{journal}{J. Phys. Chem.} \bibinfo{volume}{98} (\bibinfo{year}{1994})
  \bibinfo{pages}{11623}.
\bibitem[{Cohen and Handy(2001)}]{doi:10.1080/00268970010023435}
\bibinfo{author}{A.~J. Cohen}, \bibinfo{author}{N.~C. Handy},
  \bibinfo{journal}{Mol. Phys.} \bibinfo{volume}{99} (\bibinfo{year}{2001})
  \bibinfo{pages}{607}.
\bibitem[{Ernzerhof and Scuseria(1999)}]{ernzerhof:5029}
\bibinfo{author}{M.~Ernzerhof}, \bibinfo{author}{G.~E. Scuseria},
  \bibinfo{journal}{J. Chem. Phys.} \bibinfo{volume}{110}
  (\bibinfo{year}{1999}) \bibinfo{pages}{5029}.
\bibitem[{Bilc et~al.(2008)Bilc, Orlando, Shaltaf, Rignanese, \'I\~niguez, and
  Ghosez}]{PhysRevB.77.165107}
\bibinfo{author}{D.~I. Bilc}, \bibinfo{author}{R.~Orlando},
  \bibinfo{author}{R.~Shaltaf}, \bibinfo{author}{G.-M. Rignanese},
  \bibinfo{author}{J.~\'I\~niguez}, \bibinfo{author}{P.~Ghosez},
  \bibinfo{journal}{Phys. Rev. B} \bibinfo{volume}{77} (\bibinfo{year}{2008})
  \bibinfo{pages}{165107}.
\bibitem[{Wilson et~al.(2001)Wilson, Bradley, and Tozer}]{wilson:9233}
\bibinfo{author}{P.~J. Wilson}, \bibinfo{author}{T.~J. Bradley},
  \bibinfo{author}{D.~J. Tozer}, \bibinfo{journal}{J. Chem. Phys.}
  \bibinfo{volume}{115} (\bibinfo{year}{2001}) \bibinfo{pages}{9233}.
\bibitem[{Boese and Martin(2004)}]{boese:3405}
\bibinfo{author}{A.~D. Boese}, \bibinfo{author}{J.~M.~L. Martin},
  \bibinfo{journal}{J. Chem. Phys.} \bibinfo{volume}{121}
  (\bibinfo{year}{2004}) \bibinfo{pages}{3405}.
\bibitem[{Keal and Tozer(2005)}]{keal:121103}
\bibinfo{author}{T.~W. Keal}, \bibinfo{author}{D.~J. Tozer},
  \bibinfo{journal}{J. Chem. Phys.} \bibinfo{volume}{123}
  (\bibinfo{year}{2005}) \bibinfo{pages}{121103}.
\bibitem[{Adamo and Barone(1997)}]{Adamo1997242}
\bibinfo{author}{C.~Adamo}, \bibinfo{author}{V.~Barone},
  \bibinfo{journal}{Chem. Phys. Lett.} \bibinfo{volume}{274}
  (\bibinfo{year}{1997}) \bibinfo{pages}{242}.
\bibitem[{Zhao and Truhlar(2004)}]{doi:10.1021/jp048147q}
\bibinfo{author}{Y.~Zhao}, \bibinfo{author}{D.~G. Truhlar},
  \bibinfo{journal}{J. Phys. Chem. A} \bibinfo{volume}{108}
  (\bibinfo{year}{2004}) \bibinfo{pages}{6908}.
\bibitem[{Schmider and Becke(1998)}]{schmider:9624}
\bibinfo{author}{H.~L. Schmider}, \bibinfo{author}{A.~D. Becke},
  \bibinfo{journal}{J. Chem. Phys.} \bibinfo{volume}{108}
  (\bibinfo{year}{1998}) \bibinfo{pages}{9624}.
\bibitem[{Lynch et~al.(2000)Lynch, Fast, Harris, and
  Truhlar}]{doi:10.1021/jp000497z}
\bibinfo{author}{B.~J. Lynch}, \bibinfo{author}{P.~L. Fast},
  \bibinfo{author}{M.~Harris}, \bibinfo{author}{D.~G. Truhlar},
  \bibinfo{journal}{J. Phys. Chem. A} \bibinfo{volume}{104}
  (\bibinfo{year}{2000}) \bibinfo{pages}{4811}.

\end{thebibliography}


\end{document}